\documentclass[lettersize,journal]{IEEEtran}
\usepackage{amsmath,amsfonts, amssymb, amsthm}
\usepackage{algorithmic}
\usepackage{array}
\usepackage[caption=false,font=normalsize,labelfont=sf,textfont=sf]{subfig}
\usepackage{textcomp}
\usepackage{stfloats}
\usepackage{url}
\usepackage{verbatim}
\usepackage{graphicx}
\usepackage{cite}
\newtheorem{mydef1}{Theorem}

\newtheorem{mydef3}{Lemma}
\newtheorem{mydef4}{Corollary}
\usepackage[ruled,lined]{algorithm2e}
\newcommand\numeq[1]%
  {\stackrel{\scriptscriptstyle(\mkern-1.5mu#1\mkern-1.5mu)}{=}}
  \usepackage{breqn}
\begin{document}
	\title{Unsourced Random Access Using Multiple Stages of Orthogonal Pilots: MIMO and Single-Antenna Structures}
	\author{Mohammad Javad Ahmadi, Mohammad Kazemi, and Tolga M. Duman
 
 \thanks{This research is accepted for publications in IEEE Transactions on Wireless Communications \cite{Ahmadi2023Unsourced}, with DOI:10.1109/TWC.2023.3288376. }
 		\thanks{The authors are with the Department of Electrical and Electronics Engineering, Bilkent University, 06800 Ankara, Turkey (e-mails: \{ahmadi, kazemi, duman\}@ee.bilkent.edu.tr).}

	} 
	\maketitle
\begin{abstract}
We study the problem of unsourced random access (URA) over Rayleigh block-fading channels with a receiver equipped with multiple antennas. We propose a slotted structure with multiple stages of orthogonal pilots, each of which is randomly picked from a codebook. In the proposed signaling structure, each user encodes its message using a polar code and appends it to the selected pilot sequences to construct its transmitted signal. Accordingly, the transmitted signal is composed of multiple orthogonal pilot parts and a polar-coded part, which is sent through a randomly selected slot. The performance of the proposed scheme is further improved by randomly dividing users into different groups each having a unique interleaver-power pair. We also apply the idea of multiple stages of orthogonal pilots to the case of a single receive antenna. In all the set-ups, we use an iterative approach for decoding the transmitted messages along with a suitable successive interference cancellation technique. The use of orthogonal pilots and the slotted structure lead to improved accuracy and reduced computational complexity in the proposed set-ups, and make the implementation with short blocklengths more viable. Performance of the proposed set-ups is illustrated via extensive simulation results which show that the proposed set-ups with multiple antennas perform better than the existing MIMO URA solutions for both short and large blocklengths, and that the proposed single-antenna set-ups are superior to the existing single-antenna URA schemes.
\end{abstract}
\section{Introduction}
\indent In contrast to the conventional grant-based multiple access, where the base station (BS) waits for the preamble from devices to allocate resources to them, in grant-free random access, users transmit their data without any coordination. Removing the need for scheduling results in some benefits, such as reducing the latency and signaling overhead, which makes the grant-free set-up interesting for serving many users. Sourced and unsourced random access schemes are the main categories of grant-free random access. In the former, both the messages and identities of the users are important to the BS, so each user is assigned a unique pilot. However, this is inefficient, especially considering the next-generation wireless networks with a massive number of connected devices \cite{Shao2022Reconfigurable,Shao2020Cooperative}. In the so-called \emph{unsourced random access} (URA), which was introduced by Polyanskiy in \cite{polyanskiy2017perspective}, the BS cares only about the transmitted messages, i.e., the identity of the users is not a concern. The BS is connected to millions of cheap devices, a small fraction of which are active at a given time. In this set-up, the users employ a common codebook, and they share a short frame for transmitting their messages. In URA, the per-user probability of error (PUPE) is adopted as the performance criterion.

\indent Many low-complexity coding schemes are devised for URA over a Gaussian multiple-access channel (GMAC) including T-fold slotted ALOHA (SA)~\cite{ordentlich2017low,facenda2020efficient,vem2019user,glebov2019achievability}, sparse codes~\cite{amalladinne2018coupled,tanc2021massive,han2021sparse,ebert2021stochastic}, and random spreading~\cite{pradhan2020polar,ahmadi2021random,pradhan2021ldpc}. However, GMAC is not a fully realistic channel model for wireless communications. Therefore, in \cite{kowshik2019quasi,kowshik2020energy,kowshik2019energy,kowshik2021fundamental,andreev2020polar}, the synchronous Rayleigh quasi-static fading MAC is investigated, and the asynchronous set-up is considered in \cite{andreev2019low, amalladinne2019asynchronous}. Recently, several studies have investigated Rayleigh block-fading channels in a massive MIMO setting \cite{fengler2021non,fengler2020pilot,Gkagkos2022FASURA,ahmadi2021Unsourced}. In \cite{fengler2021non}, a covariance-based activity detection (AD) algorithm is used to detect the active messages. A pilot-based scheme is introduced in~\cite{fengler2020pilot} where non-orthogonal pilots are employed for detection and channel estimation, and a polar list decoder is used for decoding messages. Furthermore, in a scheme called FASURA \cite{Gkagkos2022FASURA}, each user transmits a signal containing a non-orthogonal pilot and a randomly spread polar code. \\
\indent The coherence blocklength is defined as the period over which the channel coefficients stay constant. As discussed in \cite{fengler2020pilot}, the coherence time can be approximated as $T_c\approx 1/4D_s$, where $D_s$ is the maximal Doppler spread. For a typical carrier frequency of $2$ GHz, the coherence time may vary in the range of $1$ ms--$45$ ms (corresponding to the transmitter speeds between $3$ km/h--$120$ km/h). Moreover, the sampling frequency should be chosen in the order of coherence bandwidth, whose typical value is between $100$ kHz--$500$ kHz in outdoor environments. Consequently, the coherence blocklength $L_c$ can range from $100$ to $20000$ samples. Although the AD algorithm in \cite{fengler2021non} performs well in the fast fading scenario (e.g., when $L_c\leq 320$), it is not implementable with larger blocklengths due to run-time complexity scaling with $L_c^2$. In contrast, the schemes in \cite{fengler2020pilot, Gkagkos2022FASURA} work well in the large-blocklength regimes (e.g., for $L_c=3200$); that is, in a slow fading environment where large blocklengths can be employed, their decoding performance is better than that of \cite{fengler2021non}. 

Most coding schemes in URA employ non-orthogonal pilots/sequences for identification and estimation purposes \cite{fengler2021non,ahmadi2021random,pradhan2021ldpc,fengler2020pilot,pradhan2020polar,Gkagkos2022FASURA}. Performance of detectors and channel estimators may be improved in terms of accuracy and computational complexity by employing a codebook of orthogonal pilots; however, this significantly increases the amount of collisions due to the limited number of available orthogonal pilot sequences. To address this problem, the proposed schemes in this paper employ multiple stages of orthogonal pilots combined with an iterative detector.

In the proposed scheme, the transmitted signal of each user is composed of $J+1$ stages: a polar codeword appended to $J$ independently generated orthogonal pilots. Thus, the scheme is called multi-stage set-up with multiple receive antennas (MS-MRA). At each iteration of MS-MRA at the receiver side, only one of the pilot parts is employed for pilot detection and channel estimation, and the polar codeword is decoded using a polar list decoder. Therefore, the transmitted pilots in the remaining $J-1$ pilot parts are still unknown. To determine the active pilots in these, we adopt two approaches. In the first one, all the pilot bits are coded jointly with the data bits and cyclic redundancy check (CRC) bits (therefore, the transmitted bits of all the pilot parts are detected after successful polar decoding). As a second approach, to avoid waste of resources, we propose an enhanced version of the MS-MRA, where only data and CRC bits are fed to the polar encoder. At the receiver side, the decoder iteratively moves through different $J+1$ parts of the signal to detect all the parts of an active user's message. Since it does not encode the pilot bits, this is called MS-MRA without pilot bits encoding (MS-MRA-WOPBE). We further improve the performance of the MS-MRA by randomly dividing users into different groups. In this scheme (called multi-stage set-up with user grouping for multiple receive antennas (MSUG-MRA)), each group is assigned a unique interleaver-power pair. Transmission with different power levels increases the decoding probability of the users with the highest power (because they are perturbed by interfering users with low power levels). Since successfully decoded signals are removed using successive interference cancellation (SIC), users with lower power levels have increased chance of being decoded in the subsequent steps. By repeating each user's signal multiple times, we further extend the idea in MS-MRA and MSUG-MRA to the case of a single receive antenna. These extensions are called multi-stage set-up with a single receive antenna (MS-SRA) and multi-stage set-up with user grouping for a single receive antenna (MSUG-SRA).\\
\indent We demonstrate that, while the covariance-based AD algorithm in \cite{fengler2021non} suffers from performance degradation with large blocklengths, and the algorithms in \cite{fengler2020pilot,Gkagkos2022FASURA} do not work well in the short blocklength regime (hence not suitable for fast fading scenarios), the MS-MRA and MSUG-MRA have a superior performance in both regimes. Furthermore, the MS-SRA and MSUG-SRA show better performance compared to similar solutions with a single receive antenna over fading channels~\cite{kowshik2019energy,andreev2019low,andreev2020polar}.\\
\indent Our contributions are as follows:
\begin{itemize}
    \item  We propose a URA set-up with multiple receive antennas, namely MS-MRA. The proposed set-up offers comparable performance with the existing schemes with large blocklengths, while having lower computational complexity. Moreover, for the short-blocklength scenario, it significantly improves the state-of-the-art.
    \item We provide a theoretical analysis to predict the error probability of the MS-MRA, taking into account all the sources of error, namely, errors resulting from pilot detection, channel estimation, channel decoding, SIC, and collisions. 
    \item We extend the MS-MRA set-up by randomly dividing the users into groups, i.e., MSUG-MRA, which is more energy-efficient than MS-MRA and other MIMO URA schemes.
    \item Two URA set-ups with a single receive antenna, called MS-SRA and MSUG-SRA, are provided by adopting the ideas of the MS-MRA and MSUG-MRA to the case of a single receive antenna. They perform better than the alternative solutions over fading channels.
\end{itemize}
\indent The rest of the paper is organized as follows. Section II presents the system model for the proposed framework. The encoding and decoding procedures of the proposed schemes are introduced in Section III. In Section IV, extensive numerical results and examples are provided. Finally, Section V provides our conclusions.\\
\indent The following notation is adopted throughout the paper. We denote the sets of real and imaginary numbers by $\mathbb{R}$ and $\mathbb{C}$, respectively. $\left[ \mathbf{T}\right]_{(l,:)}$ and $\left[ \mathbf{T}\right]_{(:,l)}$ represent the $l$th row and the $l$th column of $\mathbf{T}$, respectively; the $\mathrm{Re} \left(\mathbf{t}\right)$ and $\mathrm{Im} \left(\mathbf{t}\right)$ give the real and imaginary parts of $\mathbf{t}$, respectively; the transpose and Hermitian of matrix $\mathbf{T}$ are denoted by $\mathbf{T}^T$ and $\mathbf{T}^H$, respectively; $|.|$ denotes the cardinality of a set, $\mathbf{I}_M$ and $\mathbf{1}_{s}$ denote the $M\times M$ identity matrix and $1\times s$ all-ones vector, respectively; we use $[a_1:a_2]$ to denote $\{i\in\mathbb{Z}:a_1\leq i\leq a_2 \}$, and $\delta_{i,j}$ is the Kronecker delta.
\section{System Model}
\label{sec_sysMOdel}
We consider an unsourced random access model over a block-fading wireless channel. The BS is equipped with $M$ receive antennas connected to $K_T$ potential users, for which $K_a$ of them are active in a given frame. Assuming that the channel coherence time is larger than $L$, we divide the length-$n$ time-frame into $S$ slots of length $L$ each ($n = SL$). Each active user randomly selects a single slot to transmit $B$ bits of information. In the absence of synchronization errors, the received signal vector corresponding to the $s$th slot at the $m$th antenna is written as
\begin{align}
\mathbf{y}_{m,s} = \sum_{i\in \mathcal{K}_s}^{ }{h_{m,i}\mathbf{x}\left(\mathbf{w}(i)\right)+{\mathbf{z}_{m,s}}} ,
\label{eqs1}
\end{align}
where $\mathbf{y}_{m,s}\in \mathbb{C}^{1\times L}$, $\mathcal{K}_s$ denotes the set of active user indices available in the $s$th slot, $K_s:=|\mathcal{K}_s|$, $\mathbf{x}\left(\mathbf{w}(i)\right)\in \mathbb{C}^{1\times L}$ is the encoded and modulated signal  corresponding to the message bit sequence $\mathbf{w}(i)\in \{0,1\}^B$ of the $i$th user, $h_{m,i}\sim \mathcal{CN}(0, 1)$ is the channel coefficient between the $i$th user and the $m$th receive antenna, and ${\mathbf{z}_{m,s} }\sim \mathcal{CN}(\mathbf{0}, \sigma_z^2\mathbf{I}_L)$ is the circularly symmetric complex white Gaussian noise vector. Letting $\mathcal{K}_a$ and $\mathcal{L}_d$ be the set of active user indices and  the list of decoded messages, respectively, the PUPE of the system is defined in terms of the probability of false-alarm, $p_{fa}$, and the probability of missed-detection, $p_{md}$, as
\begin{align}
	P_e = p_{fa}+p_{md},
\end{align}
\noindent where $p_{md} =\dfrac{1}{K_a}\sum_{i\in\mathcal{K}_a}^{}{ \mathrm{Pr}(\mathbf{w}(i)\notin \mathcal{L}_d)}$ and $  p_{fa} = \mathbb{E}\left\{\dfrac{n_{fa}}{|\mathcal{L}_d|}\right\}$, with $n_{fa}$ being the number of decoded messages that were indeed not sent. The energy-per-bit of the set-up can be written as $\dfrac{E_b}{N_0}=\dfrac{LP}{\sigma_z^2 B}$, where $P$ denotes the average power of each user per channel use. The objective is to minimize the required energy-per-bit for a target PUPE. 
\section{URA with Multiple Stages of Orthogonal Pilots}
\label{sec_III_MSMRA}
\subsection{MS-MRA Encoder}
\label{sec3-a}
\begin{figure}[t!]
		\centering
		\includegraphics[width=1.01\linewidth]{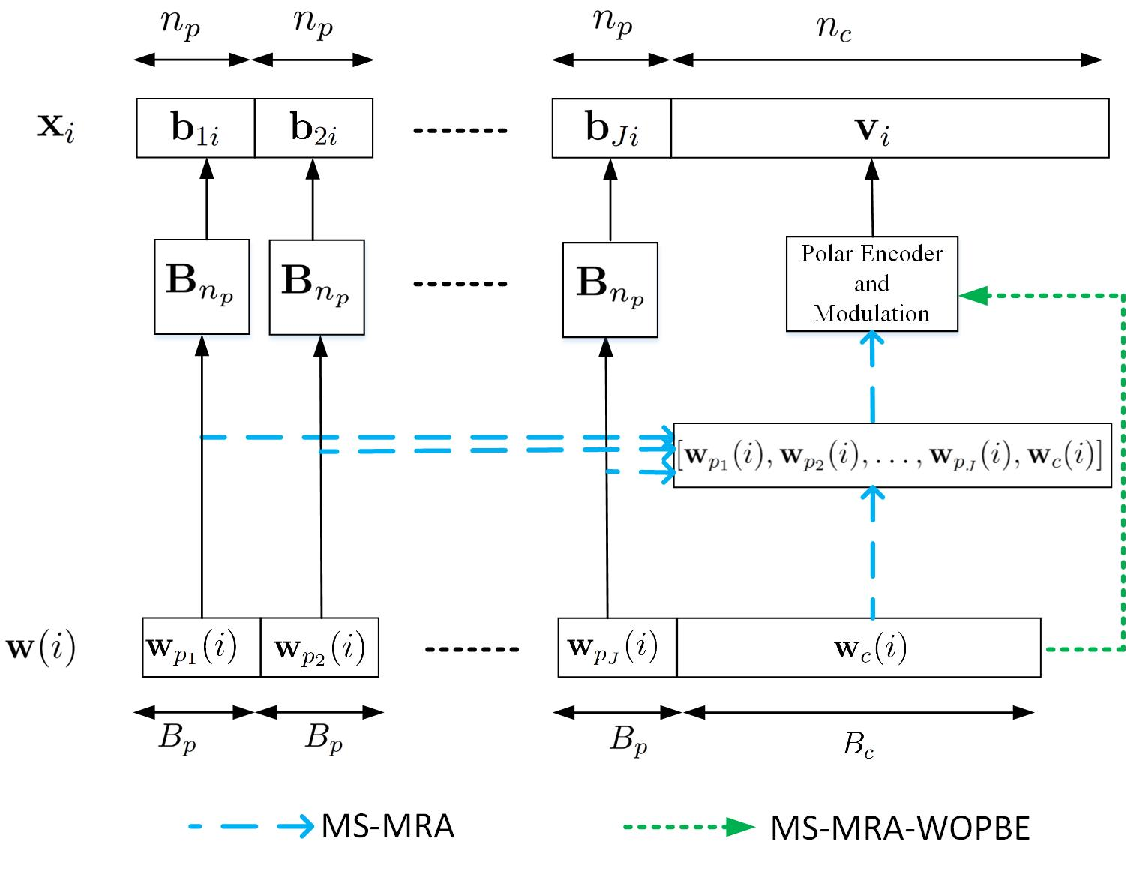}
		\caption{{\small Illustration of the encoding process in the proposed MS-MRA schemes.}	}	\label{Fig_transmission}
\end{figure}
In this part, we introduce a multi-stage signal structure which is used in both of the proposed URA set-ups. As shown in Fig. \ref{Fig_transmission}, we divide the message of the $i$th user into $J+1$ parts (one coded part and $J$ pilot parts) denoted by $\mathbf{w}_{c}(i)$ and $\mathbf{w}_{p_j}(i), j = 1,2,...,J$ with lengths $B_c$ and $B_{p}$, respectively, where $B_c+JB_{p} = B$. The $i$th user obtains its $j$th pilot sequence, $\mathbf{b}_{ji}$, with length $n_{p} = 2^{B_{p}}$ by mapping $\mathbf{w}_{p_j}(i)$ to the orthogonal rows of an $n_{p}\times n_{p}$ Hadamard matrix $\mathbf{B}_{n_{p}}$, which is generated as
\begin{align}
	\nonumber
	    \mathbf{B}_2  = \begin{bmatrix}1&1\\1&-1\end{bmatrix}, \ \ \ \
	    \mathbf{B}_{2^i} = \mathbf{B}_2 \otimes \mathbf{B}_{2^{i-1}} \ \ \forall \  \ i = 2,3, \hdots, 
\end{align}
\noindent where $\otimes$ represents the Kronecker product. Since the number of possible pilots in the orthogonal Hadamard codebook is limited, it is likely that the users will be in collision in certain pilot segments, that is, they share the same pilots with the other users. However, the parameters are chosen such that two different users are in a complete collision in all the pilot parts with a very low probability. To construct the coded sequence of the $i$th user, we accumulate all the message bits in a row vector as
\begin{align}
 \mathbf{w}(i) = \left[\mathbf{w}_{p_1}(i), \mathbf{w}_{p_2}(i), \hdots, \mathbf{w}_{p_J}(i) ,\mathbf{w}_{c}(i)\right],\label{eq9}
\end{align}
and pass it to an $\left(2n_c, \  B +r\right)$ polar code, where $r$ is the number of CRC bits. Note that contrary to the existing schemes in URA, we feed not only data bits but the pilot bit sequences to the encoder. Hence, in the case of successful decoding, all the pilot sequences for the user can be retrieved. The polar codeword is then modulated using quadrature phase shift keying (QPSK), resulting in $\mathbf{v}_i\in \{ \sqrt{{P_{c}}/{2}}(\pm 1\pm j)\}^{1\times n_c }$, where $P_c$ is the average power of the polar coded part, and Gray mapping is used. The overall transmitted signal for the $i$th user consists of $J$ pilot parts and one coded part, i.e.,
	\begin{align}
	    \mathbf{x}_i = \left[\sqrt{P_{p}}\mathbf{b}_{1i},\sqrt{P_{p}}\mathbf{b}_{2i},\hdots,\sqrt{P_{p}}\mathbf{b}_{Ji},\mathbf{v}_i\right]\in \mathbb{C}^{1\times L },\label{eq10_old}
	\end{align}
\noindent where $L = n_c+Jn_{p}$ and $P_p$ denotes the average power of the pilot sequence. Accordingly, the received signal in a slot is composed of $J+1$ parts, for which, at each iteration, the decoding is done by employing one of the $J$ pilot parts (sequentially) and the coded part of the received signal. Generally, only the non-colliding users can be decoded. Some non-colliding users in the current pilot stage may experience collisions in the other pilot parts. Therefore, by successfully decoding and removing them using SIC, the collision density is reduced, and with further decoding iterations, the effects of such collisions are ameliorated.
\subsection{MS-MRA Decoder} 
\label{sec3B}
\indent We now introduce the decoding steps of MS-MRA where the transmitted signal in \eqref{eq10_old} is received by $M$ antennas through a fading channel. The $j$th pilot part and the polar coded part of the received signal in the $s$th slot of the MS-MRA can be modeled using \eqref{eqs1} as 
	\begin{align}
\mathbf{Y}_{p_j} &= \sqrt{P_{p}} \mathbf{H}\mathbf{B}_j+\mathbf{Z}_{p_j} \in \mathbb{C}^{M\times n_{p}} , \ j = 1,2,\hdots , J,\label{eq10Pilots}\\
\mathbf{Y}_c &=  \mathbf{H}\mathbf{V}+\mathbf{Z}_c \in \mathbb{C}^{M\times n_c},\label{eq10Polar}
	\end{align}
	where $\mathbf{H}\in \mathbb{C}^{M \times K_s}$ is the channel coefficient matrix with $h_{m,i}$ in its $m$th row and $i$th column, $\mathbf{Z}_{p_j}$ and $\mathbf{Z}_c$ consist of independent and identically distributed (i.i.d.) noise samples drawn from $\mathcal{CN}(0,\sigma_z^2)$ (i.e., a circularly symmetric complex Gaussian distribution), and $\mathbf{b}_{ji}$ and $\mathbf{v}_i$ determine the rows of $\mathbf{B}_j \in \{\pm 1\}^{K_s \times n_{p}}$ and $\mathbf{V}\in \{ \sqrt{{P_{c}}/{2}}(\pm 1\pm j)\}^{K_s \times n_c}$, respectively, with $i\in \mathcal{K}_s$. Note that we have removed the slot indices from the above matrices to simplify the notation. \\
\indent The decoding process is comprised of five different steps that work in tandem. A pilot detector based on a Neyman-Pearson (NP) test identifies the active pilots in the current pilot part; channel coefficients corresponding to the detected pilots are estimated using a  channel estimator; maximum-ratio  combining (MRC) is used to produce a soft estimate of the modulated signal; after demodulation, the signal is passed to a polar list decoder; and, the successfully decoded codewords are added to the list of successfully decoded signals before being subtracted from the received signal via SIC. The process is repeated until there are no successfully decoded users in $J$ consecutive SIC iterations. In the following, $\mathbf{Y}^\prime_{p_j}$ and $\mathbf{Y}^\prime_{c}$ denote the received signals in \eqref{eq10Pilots} and \eqref{eq10Polar} after removing the list of messages successfully decoded in the current slot up to the current iteration.
\subsubsection{Pilot Detection Based on NP Hypothesis Testing}
\label{Sec_pilotDetection}
At the $j$th pilot part, we can write the following binary hypothesis testing problem:
\begin{align}
\nonumber
     &\mathbf{u}_{ji}|\mathcal{H}_0   \sim \mathcal{CN}\left(\mathbf{0},\sigma_z^2\mathbf{I}_M\right),
     \\
    & \mathbf{u}_{ji}  |\mathcal{H}_1 \sim \mathcal{CN}\left(\mathbf{0},\sigma^2_1\mathbf{I}_M\right),\label{hypothesis}
\end{align}
where $\sigma_1=\sqrt{\sigma_z^2+m_{ij} n_{p}P_{p}}$, $\mathbf{u}_{ji} := \mathbf{Y^\prime}_{p_j}\mathbf{\bar{b}}_{i}^H /\sqrt{n_{p}}$, with $\mathbf{\bar{b}}_{i}=\left[\mathbf{B}_{n_{p}}\right]_{(i,:)}$, $\mathcal{H}_1$ and $\mathcal{H}_0$ are alternative and null hypotheses that show the existence and absence of the pilot $\mathbf{\bar{b}}_{i}$ at the $j$th pilot part, respectively, and $m_{ij}$ is the number of users that pick the pilot $\mathbf{\bar{b}}_{i}$ as their $j$th pilots. 
\begin{mydef3}
\label{Theorem1}
(\hspace{-1pt}\cite[Appendix A]{ahmadi2021Unsourced}): Let $\hat{\mathcal{D}}_j$ be the estimate of the set of active rows of $\mathbf{B}_{n_{p}}$ in the $j$th pilot part. Using a $\gamma-$level Neyman-Pearson hypothesis testing (where $\gamma$ is the bound on the false-alarm probability), $\hat{\mathcal{D}}_j$ can be obtained as
\begin{align}
    \hat{\mathcal{D}}_j = \left\{l:\mathbf{u}_{jl}^H\mathbf{u}_{jl} \geq \tau_0^\prime\right \}\label{eq13},
\end{align}
where $\tau_0^\prime= 0.5\sigma_z^2\Gamma^{-1}_{2M}(1-\gamma)$, $\Gamma_k(.)$ denotes the cumulative distribution function of the chi-squared distribution with $k$ degrees of freedom $\chi^2_{k}$, and $\Gamma^{-1}_k(.)$ is its inverse.
\end{mydef3}
\noindent The detection probability of a non-colliding user ($m_{ij}=1$) is then obtained as
\begin{align}
\nonumber
 P_{D}(\delta_{NP}) =& \mathbb{P}\left(\mathbf{u}_{ji}^H\mathbf{u}_{ji} \geq \tau_0^\prime|\mathcal{H}_1 \right)\\\nonumber
\numeq{a}& 1-\Gamma_{2M}\left(\dfrac{2\tau_0^\prime}{\sigma_z^2+n_{p}P_{p}} \right) \\
= & 1-\Gamma_{2M}\left(\dfrac{\sigma_z^2\Gamma_{2M}^{-1}(1-\gamma)}{\sigma_z^2+n_{p}P_{p}} \right) ,\label{pdetect}
\end{align}
where in (a), we use the fact that $\dfrac{2}{\sigma_1^2}\mathbf{u}_{ji}^H\mathbf{u}_{ji} |\mathcal{H}_1  \sim \chi^2_{2M}$. Note that a higher probability of detection is obtained in the general case of $m_{ij}> 1 $. It is clear that the probability of detection is increased by increasing the parameters $\gamma$, $n_{p}$, $P_{p}$, and $M$.
\subsubsection{Channel Estimation}
 Let $\mathbf{B}_{\hat{\mathcal{D}}_j}\in \{\pm 1\}^{|\hat{\mathcal{D}}_j|\times n_{p}}$ be a sub-matrix of $\mathbf{B}_{n_{p}}$ consisting of the detected pilots in \eqref{eq13}, and suppose that $\tilde{\mathbf{b}}_{jk}=\left[\mathbf{B}_{\hat{\mathcal{D}}_j}\right]_{(k,:)}$ is the corresponding pilot of the $i$th user. Since the rows of the codebook are orthogonal to each other, the channel coefficient vector of the $i$th user can be estimated as
 \begin{align}
     \hat{\mathbf{h}}_{i}  =\dfrac{1}{ n_{p}\sqrt{P_{p}}}\mathbf{Y^\prime}_{p_j}\tilde{\mathbf{b}}_{jk}^T\label{eq10_Chestimate}.
\end{align} 
\indent If the $i$th user is in a collision ( $m_{ij}>1$), \eqref{eq10_Chestimate} gives an unreliable estimate of the channel coefficient vector. However, this is not important since a CRC check is employed after decoding and such errors do not propagate.
\subsubsection{MRC, Demodulation, and Channel Decoding}
Let $\mathbf{h}_{i}$ be the channel coefficient vector of the $i$th user, where $i\in \tilde{\mathcal{S}}_s$ with  $\tilde{\mathcal{S}}_s$ denoting the set of remaining users in the $s$th slot. Using $\hat{\mathbf{h}}_{i}$ in \eqref{eq10_Chestimate}, the modulated signal of the $i$th user can be estimated employing the MRC technique as
\begin{align}
    \hat{\mathbf{v}}_{i} = \hat{\mathbf{h}}_{i}^H\mathbf{Y^\prime}_c. \label{eq17}
\end{align}
 Plugging \eqref{eq10Polar} into \eqref{eq17}, $  \hat{\mathbf{v}}_{i}$ is written as
\begin{align}
     \hat{\mathbf{v}}_{i}=\hat{\mathbf{h}}_{i}^H\mathbf{h}_i\mathbf{v}_i +\mathbf{n}_i\label{eq23},
\end{align}
where $\mathbf{n}_i  =  \sum_{k\in\tilde{\mathcal{S}}_s, k\neq i}\hat{\mathbf{h}}_{i}^H\mathbf{h}_k\mathbf{v}_k+\hat{\mathbf{h}}_{i}^H\mathbf{Z}_c$. The first and second terms on the right-hand side of \eqref{eq23} are the signal and interference-plus-noise terms, respectively. We can approximate $\mathbf{n}_i$ to be Gaussian distributed, i.e., $\mathbf{n}_i \sim \mathcal{CN}(\mathbf{0}, \sigma_{oi}^2\mathbf{I}_{n_c})$, where  $\sigma_{oi}^2=\dfrac{1}{n_c}\mathbb{E}\{\mathbf{n}_i\mathbf{n}_i^H\}    = P_c\sum_{k\in\hat{\mathcal{D}}_j, k\neq i}^{} |\hat{\mathbf{h}}_{i}^H\mathbf{h}_{k}|^2 +\sigma_z^2 \|\hat{\mathbf{h}}_{i}\|^2$, which is obtained by treating the coded data sequences of different users to be uncorrelated. The demodulated signal can be obtained as
\begin{align}
\mathbf{g}_i = \left[\mathrm{Im}\left(\vartheta_{1i}\right),\mathrm{Re}\left(\vartheta_{1i}\right),\hdots,\mathrm{Im}\left(\vartheta_{n_ci}\right),\mathrm{Re}\left(\vartheta_{n_ci}\right)\right] \label{eq_23+1},
\end{align}
where $\vartheta_{ti} = \left[\hat{\mathbf{v}}_{i}\right]_{(:,t)}$. From \eqref{eq23} and \eqref{eq_23+1}, and using $\hat{\mathbf{h}}_{i}^H\mathbf{h}_i \approx \| \hat{\mathbf{h}}_{i}\|^2$, each sample of $\mathbf{g}_i$ can be approximated as $\pm \sqrt{{P_{c}}/{2}}\|\hat{\mathbf{h}}_{i}\|^2+n^\prime$, where $n^\prime\sim  \mathcal{CN}\left(0, \dfrac{\sigma_{oi}^2}{2}\right)$. The following log-likelihood ratio (LLR) is obtained as the input to the polar list decoder
\begin{align}
\mathbf{f}_i = \dfrac{2\sqrt{2P_c} \|\hat{\mathbf{h}}_{i}\|^2}{\hat{\sigma}_{oi}^2}\mathbf{g}_i \label{eq50LLR},
\end{align}
where $\hat{\sigma}_{oi}^2$ is an approximation of $\sigma_{oi}^2$ which is obtained by replacing $\mathbf{h}_k$'s by their estimates. At the $j$th pilot part, the $i$th user is declared as successfully decoded if 1) its decoder output satisfies the CRC check, and 2) by mapping the $j$th pilot part of its decoded message to the Hadamard codebook, $\tilde{\mathbf{b}}_{jk}$ is obtained. Then, all the successfully decoded messages (in the current and previous iterations) are accumulated in the set $\mathcal{S}_s$, where $|\mathcal{S}_s|+|\tilde{\mathcal{S}}_s|=K_s$. 

\subsubsection{SIC}
\begin{figure*}
		\centering
		\includegraphics[width=.94\linewidth]{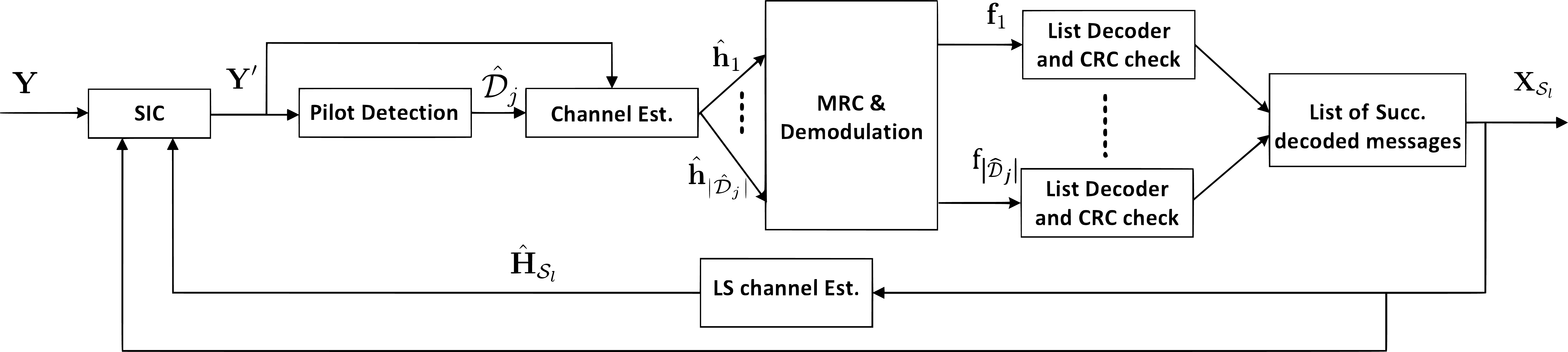}
		\caption{{\small The decoding process of MS-MRA at the $j$th pilot part and the $s$th slot.}	}	\label{Fig_decoder}
\end{figure*}

we can see in \eqref{eq9} that the successfully decoded messages contain bit sequences of pilot parts and the coded part ($\mathbf{w}_{p_j}(i), j = 1,2,...,J$ and $\mathbf{w}_{c}(i)$). Having the bit sequences of successfully decoded messages, we can construct the corresponding transmitted signals using \eqref{eq10_old}. The received signal matrix can be written as
\begin{align}
   \mathbf{Y} =  \mathbf{H}_{\mathcal{S}_s}\mathbf{X}_{\mathcal{S}_s}+\mathbf{H}_{\tilde{\mathcal{S}}_s}\mathbf{X}_{\tilde{\mathcal{S}}_s}+\mathbf{Z}_s, \label{eq21}
\end{align}
  where $\mathbf{Y}$ is obtained by merging received signal matrices of different parts, i.e., $$\mathbf{Y} = \left[\mathbf{Y}_{p_1},\hdots,\mathbf{Y}_{p_J}, \mathbf{Y}_c \right]\in \mathbb{C}^{M\times L}$$ with $\mathbf{X}_{\mathcal{S}_s}\in \mathbb{C}^{|\mathcal{S}_s|\times L }$ and $\mathbf{X}_{\tilde{\mathcal{S}}_s}\in \mathbb{C}^{|\tilde{\mathcal{S}}_s|\times L }$ including the signals in the sets $\mathcal{S}_s$ and $\tilde{\mathcal{S}}_s$, and $\mathbf{H}_{\mathcal{S}_s}\in \mathbb{C}^{M \times |\mathcal{S}_s|}$ and $\mathbf{H}_{\tilde{\mathcal{S}}_s}\in \mathbb{C}^{M \times |\tilde{\mathcal{S}_s}|}$ comprising the channel coefficients corresponding to the users in the sets $\mathcal{S}_s$ and $\tilde{\mathcal{S}}_s$, respectively. Employing the least squares (LS) technique, $\mathbf{H}_{\mathcal{S}_s}$ is estimated as
\begin{align}
    \hat{\mathbf{H}}_{\mathcal{S}_s} =  \mathbf{Y} \mathbf{X}_{\mathcal{S}_s}^H(\mathbf{X}_{\mathcal{S}_s}\mathbf{X}_{\mathcal{S}_s}^H)^{-1}.\label{ch_esimation_sic}
\end{align}
Note that $\mathbf{X}_{\mathcal{S}_s}$ consists of all the successfully decoded signals in the $s$th slot so far, and $\mathbf{Y}$ is the initially received signal matrix (not the output of the latest SIC iteration). The SIC procedure is performed as follows
\begin{align}
    \mathbf{Y^\prime}=\left[\mathbf{Y^\prime}_{p_1}, \mathbf{Y^\prime}_{p_2},\hdots,\mathbf{Y^\prime}_{p_J}, \mathbf{Y^\prime}_c \right] = \mathbf{Y}- \hat{\mathbf{H}}_{\mathcal{S}_s} \mathbf{X}_{\mathcal{S}_s}.\label{eq16}
\end{align}
Finally, $\mathbf{Y}^\prime$ is fed back to the pilot detection algorithm for the next iteration, where the next pilot part is employed. We note that if no user is successfully decoded in $J$ consecutive iterations (corresponding to $J$ different pilot parts), the algorithm is stopped. The details of the decoding stages of MS-MRA are shown in Fig. \ref{Fig_decoder} and Algorithm 1. Note that we will discuss MS-MRA-WOPBE, which deviates from the above model, in Section \ref{Sec.MS-MRA-WOPBE}.
 \begin{mydef1}
\label{Theorem_SINR}
The signal-to-interference-plus-noise ratio (SINR) at the output of MRC for a non-colliding user in the $s$th slot can be approximated as 
\begin{align}
   \beta_{s}   \approx 
   \dfrac{ \omega_{c_s}P_c \left(\omega_{p_s}\mathbb{E}\{\|\mathbf{h}_i\|^4\} +\dfrac{\sigma_z^2}{ n_{p}P_{p}} \mathbb{E}\{\|\mathbf{h}_i\|^2\}\right)}{\left(P_c   (|\tilde{\mathcal{S}}_s|-1) + \sigma_z^2\right)\left(\omega_{p_s} \mathbb{E}\{\|\mathbf{h}_i\|^2\}+\dfrac{M  \sigma_z^2}{ n_{p}P_{p} }\right)},
\label{sinr_initial}
\end{align} 
where $\omega_{p_s}=\omega_{c_s}=1-\dfrac{| \mathcal{S}_s|}{L}$ if the transmitted signals are randomly interleaved, and $\omega_{p_s}=1-\dfrac{1}{E_x}P_p| \mathcal{S}_s|$, $\omega_{c_s}=1-\dfrac{1}{E_x}P_c| \mathcal{S}_s|$, otherwise, with $E_x=Jn_pP_p+n_cP_c$.
\end{mydef1}
\begin{proof}
See Appendix \ref{Appendix_ch_dec}.
\end{proof}
We employ the above approximate SINR expression 1) to estimate the error probability of MS-MRA analytically, and 2) to determine the optimal power allocation for each group in MSUG-MRA. We further note that using this SINR approximation, the performance of the MS-MRA is well predicted in the low and medium $K_a$ regimes (see Fig. \ref{fig_theory}). The reason why the SINR approximation does not work well in the high $K_a$ regime is the employed approximations in Lemma \ref{lemma_y_cy_p} (see the Appendix \ref{Appendix_ch_dec} for details).
\begin{algorithm}
	{\small
		\caption{ The proposed MS-MRA decoder.} 
	\For( Different slots){{$l = 0,1, \hdots, S $}}
		{
		$\mathcal{S}_s$    $\ \ =\emptyset$.\\
		$\mathrm{flag} \ = 1$.\\
					$t^\prime = 0$ (\textit{$t^\prime$ denotes the iteration index}).\\
			\While( {}){$\mathrm{flag}=1$}
			{
			$t^\prime \ \ = t^\prime+1.$\\

				\For(different pilot parts ){{$j = 1, 2,...,J $}}
				{
					 \textbf{Pilot detection}:  estimate $\hat{\mathcal{D}}_j$ using \eqref{eq13}.\\
					 \textbf{Ch. estimation}:  estimate channel coefficient using \eqref{eq10_Chestimate}.\\
\If{MS-MRA}{
 \For(different detected pilots){{$i \in  \hat{\mathcal{D}}_j$}}
				{
  \textbf{MRC estimation}: obtain $\hat{\mathbf{v}}_{i}$ using \eqref{eq17}.\\
  \textbf{Demodulation}: obtain $\mathbf{g}_i$ using \eqref{eq_23+1}.\\
  \textbf{Decoding}:  \footnotesize{pass $\mathbf{f}_i$ in \eqref{eq50LLR} to list decoder.}\\
				}
}

\If{MS-MRA-WOPBE}{
Perform IISD in Section \ref{Sec.MS-MRA-WOPBE_decoder}.
}
$\mathcal{S}_{t^\prime j}$: set of successfully decoded users in the current iteration.\\
						$\mathcal{S}_s = 	\mathcal{S}_s \bigcup \mathcal{S}_{t^\prime j} $.\\
      			\textbf{LS-based ch. estimation}:  estimate $ \hat{\mathbf{H}}_{\mathcal{S}_s}$ using \eqref{ch_esimation_sic}.\\
					\textbf{SIC}:  update $\mathbf{Y^\prime}_{p_j}$ and $\mathbf{Y^\prime}_{c}$ using \eqref{eq16}.    
				}
				\If{$\bigcup_{j=1}^J \mathcal{S}_{t^\prime j}=\emptyset$}{$\mathrm{flag}=0$.}
			}
			}
			}
	\end{algorithm}
\subsection{Analysis of MS-MRA}
In this part, the PUPE of the MS-MRA is analytically calculated, where errors resulting from the collision, pilot detection, and polar decoder are considered. For our analyses, we assume that after successfully decoding and removing a user using a pilot part, the decoder moves to the next pilot part. Hence, in the $t$th iteration of the $s$th slot, we have 
\begin{subequations}
\label{assumption1}
\begin{align}
|\mathcal{S}_s|&=t-1,\\
|\tilde{\mathcal{S}}_s|&=K_s-t+1.
\end{align}
\end{subequations}
\begin{mydef3}
\label{lemma1_1}
Let $\xi_{k}$ be the event that $k$ out of $K_s$ users remain in the $s$th slot, and define $\eta_i:=\|\mathbf{h}_i\|^2$, where $\mathbf{h}_i\sim \mathcal{CN}(\mathbf{0},\mathbf{I}_M)$. Assuming that the strongest users with highest $\eta_i$ values are decoded first, we have
\begin{align}
    \mathbb{E}\{\eta_i^m|\xi_{k}\}=\mu_{(k,m)},
\end{align}
where $\mu_{(k,m)}:=  \dfrac{\int_{-\infty}^{\bar{x}_{k} } \eta^m f_{2M}^{\chi^2}(2\eta)d\eta }{ \int_{-\infty }^{\bar{x}_{k}}f_{2M}^{\chi^2}(2\eta) d\eta}$, with $f_{k}^{\chi^2}(.)$ denoting the PDF of the chi-squared distribution with $k$ degrees of freedom and $\bar{x}_k=0.5 \Gamma_{2M}^{-1}(k/K_s)$.
\end{mydef3}
\begin{proof}
In the first iteration of the $s$th slot for which no user is decoded yet (all the $K_s$ active users are available), since $\mathbf{h}_i\sim \mathcal{CN}(\mathbf{0},\mathbf{I}_M)$, we have $2\eta_i|\xi_{K_s}   \sim \chi^2_{2M}$. We assume that the users with higher values of $\eta_i$ are decoded first. Hence, if in an iteration, $k$ out of $K_s$ users remain in the slot, the distribution of $\eta_i$ is obtained by $2\eta_i|\xi_{k}\sim\left\{\chi^2_{2M}\right\}_{k/{K_s}},$ where $\left\{.\right\}_{\beta}$ removes $1-\beta$ portion of the samples with higher values from the distribution and normalizes the distribution of the remaining samples, i.e., \[\mathbb{P}(\eta_i=y|\xi_{k})= \dfrac{ f_{2M}^{\chi^2}(2y) }{ \int_{-\infty }^{\bar{x}_{k}}f_{2M}^{\chi^2}(2y) dy}, y<\bar{x}_{k},\] where $\bar{x}_{k}$ is obtained by solving the following equation $\mathbb{P}\left(\eta_i<\bar{x}_{k} |\xi_{K_s}\right)=k/{K_s}$, which results in $\bar{x}_{k}=0.5 \Gamma_{2M}^{-1}(k/K_s)$. Therefore, we obtain 
\begin{align}
    \mathbb{E}\{\eta_i^m|\xi_{k}\}=   \dfrac{\int_{-\infty}^{\bar{x}_{k} } \eta^m f_{2M}^{\chi^2}(2\eta)d\eta }{ \int_{-\infty }^{\bar{x}_{k}}f_{2M}^{\chi^2}(2\eta) d\eta}.
\end{align}
\end{proof}
\indent We can see from \eqref{eq_23+1} and \eqref{eq50LLR} that the input of the polar decoder is a $1\times 2n_c$ real codeword. Thus, the average decoding error probability of a non-colliding user in the $t$th iteration of a slot with $K_s$ users can be approximated as (see \cite{Polyanskiy2010Channel})
\begin{align}
  P_{ K_s ,t}^{dec} \approx  Q\left(\dfrac{ 0.5\log\left(1+\alpha_{ K_s ,t} \right)-\dfrac{B+r}{2n_c}}{ \sqrt{ \dfrac{1}{2n_c}\dfrac{\alpha_{ K_s ,t}(\alpha_{ K_s ,t}+2)\log^2e}{2(\alpha_{ K_s ,t}+1)^2}}}\right), \label{eq_error_decoding}
\end{align}
where $Q(.)$ denotes the standard $Q$-function, and $\alpha_{ K_s ,t}$ is the SINR of a non-colliding user in the $t$th iteration of a slot with $K_s$ users, which is calculated using Theorem \ref{Theorem_SINR}, Lemma \ref{lemma1_1}, and \eqref{assumption1} as
\begin{align}
   \alpha_{ K_s ,t} \approx 
   \dfrac{ s_{c_t}P_c \left(s_{p_t}\mu_{(K_s-t+1,2)} +\dfrac{\sigma_z^2}{ n_{p}P_{p}} \mu_{(K_s-t+1,1)}\right)}{\left(P_c   (K_s-t) + \sigma_z^2\right)\left(s_{p_t} \mu_{(K_s-t+1,1)}+\dfrac{M  \sigma_z^2}{ n_{p}P_{p} }\right)},
\label{eq_prob_dec}
\end{align}
where $s_{p_t} = 1-P_p\dfrac{t-1}{E_x}$ and $s_{c_t}  = 1-P_c\dfrac{t-1}{E_x}$. Note that since the powers of signal and  interference-plus-noise terms of $\hat{\mathbf{v}}_{i}$ are equal in their real and imaginary parts, the SINRs of $\mathbf{f}_i$ in \eqref{eq50LLR} and $\hat{\mathbf{v}}_{i}$ are the same. Therefore, in \eqref{eq_error_decoding}, we employ the SINR calculated in Theorem \ref{Theorem_SINR} for the input of the polar list decoder.\\
\indent Since decoding in the initial iterations well represents the overall decoding performance of the MS-MRA, we approximate the SINR of the first iteration by setting $t=1$ in \eqref{eq_prob_dec} as
\begin{align}
   \alpha_{K_s,1}\approx \dfrac{P_cM}{ \left(\sigma_z^2+P_cK_s\right)\left(1+\dfrac{\sigma_z^2}{ n_{p}P_{p}}\right)}.\label{eq_simplified_SINR}
\end{align}
Concentrating on \eqref{eq_error_decoding}, we notice that $P_{K_s,1}^{dec}$ is a decreasing function of $n_c$ and $\alpha_{K_s,1}$. Besides, \eqref{eq_simplified_SINR} shows that $\alpha_{K_s,1}$ increases by decreasing $n_c$ and $J$ (considering $K_s\approx K_a(Jn_p+n_c)/n$), and increasing $M$, $P_c$, and $P_p$, however, it is not a strictly monotonic function of $n_p$. Since our goal is to achieve the lowest $P_{K_s,t}^{dec}$ by spending the minimum ${E_b}/{N_0}=(n_cP_c+Jn_pP_p)/B$, we can optimize the parameters $n_c$, $n_p$, $P_c$, and $P_p$.
\begin{mydef1}
\label{Thm_collision}
In the $t$th iteration of the $s$th slot, the probability of collision for a remaining user $i\in \tilde{\mathcal{S}}_s$ can be approximated as
\begin{align}
    P_{K_s,t}^{col} \approx 1- \dfrac{N_1^{\left(t\right)}}{K_s-t+1},\label{collisionProb}
\end{align}
where $N_i^{(k)}$ denotes the average number of pilots that are in $i$-collision (selected by $i$ different users) in the $k$th iteration, which is calculated as
\begin{align}
N_i^{(k+1)} \approx N_i^{(k)}+\begin{cases} \kappa_k\left((i+1)N_{i+1}^{(k)}-iN_i^{(k)}\right) & i\geq 2\\\kappa_k\left(2N_{2}^{(k)}-N_1^{(k)}\right) -\dfrac{1}{J}& i = 1\end{cases}, \label{eq_Recers_Ni}
\end{align}
where $\kappa_k = \dfrac{J-1}{J(K_s-k+1)}$, and $N_i^{(1)}\approx n_p f_p(i;K_s/n_p)$ with $f_p(i;a)$ denoting the probability mass function (PMF) of the Poisson distribution with the parameter $a$.
\end{mydef1}
\begin{proof}
See Appendix \ref{Appendix_collision}.
\end{proof}
Note that to extend the result in Theorem \ref{Thm_collision} to an SIC-based system with only one pilot sequence (orthogonal or non-orthogonal), we only need to set $J=1$ in the above expressions.
\noindent From \eqref{collisionProb}, the collision probability in the first iteration can be calculated as $P_{K_s,1}^{col} \approx 1- e^{-K_s/n_p}$, which is a decreasing function of $n_p$. Since the overall decoding performance of the system depends dramatically on the collision probability in the first iteration, we can increase $n_p$, however, this results in additional overhead. 
\begin{mydef4}
\label{Lemma2}
Assuming a relatively large CRC length (hence negligible $p_{fa}$), the PUPE of the MS-MRA with $S$ slots and $K_a$ active users can be approximated as
\begin{align}
  P_e \approx 1-\sum_{r=1}^{K_a}(1-\epsilon_r)  \binom{K_a-1}{r-1}\left(\dfrac{1}{S}\right)^{r-1}\left(1-\dfrac{1}{S}\right)^{K_a-r},\label{eq18Ach}
\end{align}
where $\epsilon_r$ denotes the PUPE of a slot with $r$ users, which is obtained as
\begin{align}
    \epsilon_r\approx \sum_{j=1}^{r} \dfrac{r-j+1}{r}p_{j,r},
\end{align}
with $p_{j,r} = (e_{j,r})^{r-j+1}\prod_{f=1}^{j-1}\left(1-(e_{f,r})^{{r}-f+1}\right)$, and 
\begin{align}
 e_{t,r} = 1-   P_{D}(\delta_{NP}) \left(1-  P_{r,t}^{dec} \right)\left(1-P_{r,t}^{col}\right), \label{error_iter_slot}
\end{align}
where $P_{r,t}^{dec}$, $P_{r,t}^{col}$, and $ P_{D}(\delta_{NP})$ are computed in \eqref{eq_error_decoding}, Theorem \ref{Thm_collision}, and \eqref{pdetect}, respectively.
\end{mydef4}
\noindent Note that the result in Corollary \ref{Lemma2} can also be used in any other slotted system with SIC by replacing appropriate $e_{j,r}$.
\subsection{MS-MRA-WOPBE}
\label{Sec.MS-MRA-WOPBE}
As discussed in Section \ref{sec3-a}, in the MS-MRA scheme, the pilot bits are fed to the polar encoder along with the data and CRC bits. To improve the performance by decreasing the coding rate, the MS-MRA-WOPBE scheme passes only the data and CRC bits to the encoder. To detect the bit sequences of different parts of the message, it employs an extra iterative decoding block called iterative inter-symbol decoder (IISD) (described in Section \ref{Sec.MS-MRA-WOPBE_decoder}). At each step of IISD, it detects one part of a user's signal (polar or pilot part), appends the detected part to the current pilot (which was used for channel estimation in the previous step) to have an extended pilot, and re-estimates the channel coefficients accordingly. The encoding and decoding procedures of MS-MRA-WOPBE are described below.
\subsubsection{Encoder}
The $i$th user encodes its bits using the following steps (the general construction is shown in Fig. \ref{Fig_transmission}). Similar to the MS-MRA encoder in Section \ref{sec3-a}, $B$ information bits are divided into $J+1$ parts as in \eqref{eq9}, and the transmitted signal is generated as in \eqref{eq10_old}. The only difference is in the construction of the QPSK signal. The encoder in MS-MRA-WOPBE defines two CRC bit sequences as $\mathbf{c}_2(i) = \mathbf{w}(i)\mathbf{G}_2 $ and $\mathbf{c}_1(i) = [\mathbf{w}_{c}(i),\mathbf{c}_2(i)]\mathbf{G}_1 $, where $\mathbf{G}_2 \in \{0,1\}^{B\times r_2}$ and $\mathbf{G}_1\in \{0,1\}^{(B_c+r_2)\times r_1}$ are generator matrices known by the BS and users. Then, it passes $[\mathbf{w}_{c}(i),\mathbf{c}_2(i),\mathbf{c}_1(i)]$ to an $\left(2n_c, \  B_c+r_1+r_2 \right)$ polar encoder, and modulates the output by QPSK to obtain $\mathbf{v}_i\in \{ \sqrt{{P_{c}}/{2}}(\pm 1\pm j)\}^{1\times n_c }$.
\subsubsection{Decoder}
\label{Sec.MS-MRA-WOPBE_decoder}
As shown in Algorithm 1, MS-MRA-WOPBE exploits the same decoding steps as the MS-MRA scheme, except for the IISD step. We can see in Algorithm 1 that the $j$th pilot of the $i$th user is detected before employing the IISD. Then, IISD must detect the data (polar) sequence and the $f$th pilot of the $i$th user, where $f=1,...,J, f\neq j$. In the following, IISD is described in detail.\\
    \textbf{Step 1 }[Detecting $\mathbf{w}_c(i) \forall i\in \hat{\mathcal{D}}_j$]:  We first obtain $\mathbf{g}_i$ using \eqref{eq_23+1}, where $   \hat{\mathbf{v}}_{i} = \hat{\mathbf{h}}_{i}^H\mathbf{R}_h^{-1}\mathbf{Y^\prime}_c$, and $\mathbf{R}_h=\sigma_z^2\mathbf{I}_M+P_c\sum_{l\in \hat{\mathcal{D}}_j}\hat{\mathbf{h}}_{l}\hat{\mathbf{h}}_{l}^H$. Then, we pass $\mathbf{f}_i = \dfrac{2\sqrt{2P_c}}{1-P_c\hat{\mathbf{h}}_{i}^H\mathbf{R}_h^{-1}\hat{\mathbf{h}}_{i}}\mathbf{g}_i$ to the list decoder. A CRC check $\mathrm{flag_{CRC1}}(i)\in \{0,1\}$ and an estimate of $[\mathbf{w}_{c}(i),\mathbf{c}_2(i),\mathbf{c}_1(i)]$ \footnote{In the output of the polar list decoder, there is a list of possible messages. If more than one messages satisfy the CRC check ($\mathbf{c}_1(i) = [\mathbf{w}_{c}(i),\mathbf{c}_2(i)]\mathbf{G}_1$), the most likely of them is returned as the detected message and the CRC flag is set to one. Otherwise, the most likely message is returned as the detected message and the CRC flag is set to zero.} are obtained by the polar list decoder.\\
\noindent \textbf{Step 2 }[Updating $\hat{\mathbf{h}}_{i}$]: Since the $j$th pilot and polar codeword of the $i$th user are detected so far, we append them to construct a longer signal as $\mathbf{q}_i = [\mathbf{b}_{ji}, \mathbf{v}_i]\in \mathbb{C}^{1\times (n_p+n_c)}$. Then, we update $\hat{\mathbf{h}}_{i}$ by MMSE estimation as $   \hat{\mathbf{h}}_{i} = \mathbf{Y^\prime}_q \mathbf{R}_q^{-1}   \mathbf{q}_i^H$, where $\mathbf{R}_q=\sigma_z^2\mathbf{I}_{(n_p+n_c)}+\sum_{l\in \hat{\mathcal{D}}_j}\mathbf{q}_l^H\mathbf{q}_l$, and $\mathbf{Y^\prime}_q = [\mathbf{Y^\prime}_{p_j},\mathbf{Y^\prime}_c]$.\\ \textbf{Step 3 }[Detecting $\mathbf{w}_{p_f}(i) \forall i\in \hat{\mathcal{D}}_j, f\neq j$]: Assuming that the $t$th row of the Hadamard matrix is active in the $f$th pilot part ($f\neq j$), we estimate the corresponding channel coefficient as $\mathbf{s}_{ft}=\dfrac{1}{ n_{p}\sqrt{P_{p}}}\mathbf{Y^\prime}_{p_f}\tilde{\mathbf{b}}_{ft}^T$ (see \eqref{eq10_Chestimate}). To find the $f$th pilot sequence of the $i$th user, we find the pilot whose corresponding channel coefficient vector is most similar to $\hat{\mathbf{h}}_{i}$, i.e., we maximize the correlation between $\hat{\mathbf{h}}_{i}$ and $\mathbf{s}_{ft}$ as
\begin{align}
         \hat{t}_{fi} = \max_t{\dfrac{|\hat{\mathbf{h}}_{i}^H\mathbf{s}_{ft}|^2}{\mathbf{s}_{ft}^H\mathbf{s}_{ft}}}, f=1,...,J, f\neq j. \label{eq_maximization}
\end{align}
\noindent \textbf{Step4 }[Updating $\hat{\mathbf{h}}_{i}$]:
       Since the bit sequences of all $J+1$ parts are detected, we can construct $\mathbf{x}_i$ using \eqref{eq10_old}. The channel coefficient vector can be updated by MMSE as $\hat{\mathbf{h}}_{i} = \mathbf{Y^\prime} \mathbf{R}^{-1}   \mathbf{x}_i^H$, where $\mathbf{R}=\sigma_z^2\mathbf{I}_{L}+\sum_{l\in \hat{\mathcal{D}}_j}\mathbf{x}_l^H\mathbf{x}_l$. If the number of users that satisfy $\mathrm{flag_{CRC1}}(i)=1$ is not changed in an iteration, the iteration is stopped, otherwise, the algorithm goes to Step 1 for another iteration with updated $\hat{\mathbf{h}}_{i}$. Users whose bit sequences satisfy $\mathbf{c}_2(i) = \mathbf{w}(i)\mathbf{G}_2 $ and $\mathbf{c}_1(i) = [\mathbf{w}_{c}(i),\mathbf{c}_2(i)]\mathbf{G}_1$ are added to the set $\mathcal{S}_{t^\prime j}$ as successfully decoded users of the current iteration.

\subsection{MSUG-MRA}
Different from MS-MRA where the power of every user is the same and signals are not interleaved, MSUG-MRA defines $G$ groups, each being assigned unique interleaver and power pair ($\pi_g(.),P_{p_g},P_{c_g}$), $g=1,2,...,G$. We assume that $ \phi = \dfrac{P_{p_g}}{P_{c_g}}$ is constant in all groups, hence each group can be identified with a unique interleaver-power pair $(\pi_g(.),P_{c_g})$, which is known at both transmitter and receiver sides. The details of encoding and decoding procedures as well as the power selection strategy are explained below. Note that we assume without loss of generality that $P_{c_1}<P_{c_2}...<P_{c_G}$.
\subsubsection{Encoder}
\label{Section_MSUG_encodedr}
The encoding is adopted as follows:
\begin{itemize}
    \item Every user randomly selects a group, e.g., with index $g$.
    \item Each user employs $P_{c_g}$ and $\phi P_{c_g}$ as the powers of the coded and pilot parts, with which it generates its multi-stage signal $\mathbf{x}_i$ similar to MS-MRA (according to \eqref{eq10_old}). 
   \item The transmitted signal is created as $\tilde{\mathbf{x}}_i = \pi_g(\mathbf{x}_i)$.
\end{itemize}
\subsubsection{Decoder}
\label{section_MSUGMRA_decoder}
In each iteration, the decoder tends to decode the messages belonging to the users of the dominant group (the $G$th group with the highest power level). After decoding and removing users in the $G$th group, users in the $(G-1)$st group become the dominant ones. Using the same trend, all the groups have the chance to be the dominant group at some point. Since users in different groups are interleaved differently, signals of users in other groups are uncorrelated from the signals in the dominant group. Thus, letting the $g_0$th group to be dominant, we approximately model the $f$th signal in the the $g$th group ($g\neq g_0$) as 
\begin{align}
    \tilde{\mathbf{x}}_f\sim \mathcal{CN}(\mathbf{0},\zeta P_{c_g} \mathbf{I}_L), \label{eq_apprx_gaussian}
\end{align}
where $\zeta =\dfrac{J\phi n_p+n_c}{L}$. Therefore, when the $g_0$th group is dominant (the users in the groups with indices greater than $g_0$ are already removed using SIC), users in the $g_0$th group are perturbed by i.i.d. noise samples drawn from $\mathcal{CN}(0,\delta_{g_0} )$, with $\delta_{g_0} \approx \zeta K_0\sum_{g=1}^{g_0-1}P_{c_g}+\sigma_z^2$, where $K_0=\dfrac{K_a}{SG}$ is the average number of users in each group of the current slot. Consequently, by replacing $\sigma_z^2$, $P_p$, and $P_c$ with $\delta_{g_0}$, $\phi P_{c_{g_0}}$, and $P_{c_{g_0}}$ in the decoding steps of MS-MRA (in Section \ref{sec3B}), the decoding procedure of MSUG-MRA is obtained as:
\begin{itemize}
    \item Deinterleave the rows of the received signals:$\mathbf{\tilde{Y}^\prime}_{p_j} =\pi_{g_0}^{-1}(\mathbf{Y^\prime}_{p_j})$ and $\mathbf{\tilde{Y}^\prime}_{c} =\pi_{g_0}^{-1}(\mathbf{Y^\prime}_{c})$.
    \item Find active pilots as $$ \hat{\mathcal{D}}_j = \left\{l:\mathbf{\tilde{u}}_{jl}^H\mathbf{\tilde{u}}_{jl} \geq 0.5\delta_{g_0}\Gamma^{-1}_{2M}(1-\gamma)\right \},$$ where $\mathbf{\tilde{u}}_{ji} := \mathbf{\tilde{Y}^\prime}_{p_j}\mathbf{\bar{b}}_{i}^H /\sqrt{n_{p}}$.
    \item Channel estimation and MRC: $\hat{\mathbf{v}}_{i} = \hat{\mathbf{h}}_{i}^H\mathbf{\tilde{Y}^\prime}_c$, where $     \hat{\mathbf{h}}_{i}  =\dfrac{1}{ n_{p}\sqrt{\phi P_{c_{g_0}}}}\mathbf{\tilde{Y}^\prime}_{p_j}\tilde{\mathbf{b}}_{jk}^T$, and $\tilde{\mathbf{b}}_{jk}$ is one of the detected pilots.
    \item Pass $\mathbf{f}_i = \dfrac{2\sqrt{2 P_{c_{g_0}}} \|\hat{\mathbf{h}}_{i}\|^2}{\hat{\sigma}_{oi}^2}\mathbf{g}_i$ to the polar decoder, where $\hat{\sigma}_{oi}^2=P_{c_{g_0}}\sum_{k\in\hat{\mathcal{D}}_j, k\neq i}^{} |\hat{\mathbf{h}}_{i}^H\hat{\mathbf{h}}_{k}|^2 +\delta_{g_0} \|\hat{\mathbf{h}}_{i}\|^2$, and $\mathbf{g}_i$ is defined in \eqref{eq_23+1}.
\item Regenerate signals of successfully decoded users according to Section \ref{Section_MSUG_encodedr} (using ($\pi_{g_0}(.)$,$P_{c_{g_0}}$) pair), and collect them in the rows of $\mathbf{\tilde{X}}_{\mathcal{S}_s}$. 
\item Apply LS-based SIC similar to \eqref{eq16}, i.e., $\mathbf{Y}^\prime = \mathbf{Y}(\mathbf{I}_L-  \mathbf{\tilde{X}}_{\mathcal{S}_s}^H(\mathbf{\tilde{X}}_{\mathcal{S}_s}\mathbf{\tilde{X}}_{\mathcal{S}_s}^H)^{-1} \mathbf{\tilde{X}}_{\mathcal{S}_s})$.
\end{itemize}
Note that this loop is repeated for $G$ different group indices and $J$ different pilot parts, and the iteration is stopped if there is no successfully decoded users in $GJ$ consecutive iterations.
\subsubsection{Power Calculation}
When MSUG-MRA starts the decoding in the $g_0$th group, there are $|\mathcal{S}_s| \approx K_0 (G-g_0)$ successfully decoded users from previous groups (with higher power levels), $|\tilde{\mathcal{S}}_s|=K_0$ users remain in the $g_0$th group, and users in the current group are perturbed with a complex Gaussian noise with covariance matrix $\delta_{g_0} \mathbf{I}_M$. Therefore, the SINR of a non-colliding user in the current group can be calculated by replacing $|\tilde{\mathcal{S}}_s| \approx K_0 $, $|\mathcal{S}_s|=K_0(G-g_0)$, $\mathbb{E}\{\|\mathbf{h}_i\|^2\}=M$, $\mathbb{E}\{\|\mathbf{h}_i\|^4\}=M^2$, $P_c = P_{c_{g_0}}$, $P_p = \phi P_{c_{g_0}}$, $\sigma_z^2 \approx \delta_{g_0}$, and $\omega_{p_s}=\omega_{c_s}=1-\dfrac{| \mathcal{S}_s|}{L}$ in \eqref{sinr_initial} as
\begin{align}
   \beta_{g_0}^\prime  \approx \dfrac{ \rho_{g_0}MP_{c_{g_0}}^2+\dfrac{\delta_{g_0}}{n_p \phi }P_{c_{g_0}}  }{\left(P_{c_{g_0}}   (K_0-1) + \delta_{g_0}\right)\left( P_{c_{g_0}}+\dfrac{  \delta_{g_0}}{\rho_{g_0}  n_{p}\phi  }\right)},
\end{align} 
where $\rho_{g_0} = 1-\dfrac{K_0 (G-g_0)}{L}$. To impose similar performance on different groups, we set $\beta_{1}^\prime=\beta_{2}^\prime=\hdots=\beta_{G}^\prime$. Solving this equation, the power of the $g$th group satisfies $c_1 P_g^2+c_2 P_g+c_3=0$, 
where $c_1 = (K_0-1) -\dfrac{\rho_{g}M}{\beta_{g-1}^\prime}$, $c_2 = \delta_{g}\left(1+\dfrac{(K_0-1)}{\phi   n_p \rho_g}-\dfrac{1}{\phi   n_p \beta_{g-1}^\prime}\right)$, $c_3 = \dfrac{\delta_{g}^2}{\phi   n_p \rho_g}$. Solving this equation, we have
\begin{align}
  &  P_t =  \dfrac{-c_2 + \sqrt{c_2^2 - 4c_1c_3}}{2c_1}, \\\nonumber
   & \mathrm{s.t.  \ }    \dfrac{1}{G}\sum_{f=1}^G P_f=P \ \mathrm{and}  \ \ P_t \in \mathbb{R}^+.
\end{align}
Note that the MS-MRA scheme is a special case of the MSUG-MRA with $G=1$.
 \subsection{MS-SRA and MSUG-SRA}
\label{sec3D}
In this part, we apply the proposed MIMO coding schemes to the case of a single receive antenna. To accomplish this, we repeat each user's length-$L$ signal multiple times to create temporal diversity in MS-SRA and MSUG-SRA. Accordingly, we divide the whole frame into $V$ sub-frames of length $n^\prime  = n/V$, then divide each sub-frame into $S$ slots of length $L = n^\prime/S$. Each user randomly selects a slot index, namely $s$, and transmits its signal, through the $s$th slot of each sub-frame. Assuming the coherence time to be $L$, each sub-frame is analogous to a receive antenna. Therefore, the transmitted messages in MS-SRA and MSUG-SRA can be decoded using MS-MRA and MSUG-MRA decoders in Sections \ref{sec3B} and \ref{section_MSUGMRA_decoder}, respectively, considering $V$ receive antennas. Since each user repeats its signal $V$ times, for this case, we have ${E_b}/{N_0}=\dfrac{VLP}{\sigma_z^2 B}$.
\subsection{Computational Complexity}
We focus on the number of multiplications as a measure of the computational complexity, and make a complexity comparison among the proposed and existing URA solutions. The per-iteration computational complexity of the MS-MRA in a slot is calculated as follows: The pilot detection in \eqref{eq13} has a complexity of $\mathcal{O}( n_p^2MJS)$ corresponding to $J$ different pilot parts and $S$ different slots, where $\mathcal{O}(.)$ is the standard big-O notation, denoting the order of complexity. The channel estimator in \eqref{eq10_Chestimate} does not require any extra computation, because $\hat{\mathbf{h}}_{i}$ corresponds to $\mathbf{u}_{ji}$ which is calculated before for pilot detection; the MRC in \eqref{eq17} has a complexity of $\mathcal{O}(\sum_{j=1}^{J} |\mathcal{D}_j|M n_c S )$; to compute the LLR in \eqref{eq50LLR}, the required computational complexity is $\mathcal{O}( \sum_{j=1}^{J}|\mathcal{D}_j|^2 M S )$; the computational complexity of the polar list decoder is \cite{Arikan_channl} $\mathcal{O}(\sum_{j=1}^{J}|\mathcal{D}_j|n_c\log n_c S)$; and, the SIC has a complexity of $\mathcal{O}(ML|\mathcal{S}_s|S+|\mathcal{S}_s|^2LS)$. We know from \eqref{Eq_numCol} that in the first iteration, we have $|\mathcal{D}_j|\approx    n_p-n_p e^{{-K_a}/{(n_pS)}}< {K_a}/{S}$, and $|\mathcal{S}_s|=0$; in the last iterations, we have $|\mathcal{S}_s|\approx K_a/S$ and $|\mathcal{D}_j| \approx 0$. Hence, considering $M\gg \log n_c$ and $n_c|\mathcal{D}_j|\gg n_p$, we can compute the computational complexity of the MS-MRA in the first and last iterations as $\mathcal{O}\left(K_aMJ(n_c+K_a/S)\right)$ and $\mathcal{O}\left(LK_a(M+K_a/S)\right)$, respectively. Considering the computational complexity in the intermediate iterations to be in the same order, the per-iteration computational complexity of the MS-MRA can be bounded by
$\mathcal{O}\Bigl(n_p^2MJS+\max\left(K_aMJ(n_c+K_a/S) ,LK_a(M+K_a/S)\right)\Bigl)$. Note that the computational complexity of MSUG-MRA is in the same order as MS-MRA, and for MS-SRA and MSUG-SRA schemes, the computational complexity is obtained by replacing $M$ by $V$ in the above figures. Note that by employing a low-complexity adaptive filter \cite{Ahmadi2020Efficient,Abadi2019Diffusion, Abadi2019Two}, we can considerably reduce the computational complexity of the LS-based channel estimator in \eqref{ch_esimation_sic} (hence the total computational complexity of the proposed schemes).


Looking at Algorithm 1, we can infer that MS-MRA-WOPBE is obtained by employing the same pilot detector (with complexity $\mathcal{O}( n_p^2MJS)$), channel estimator (does not incur any extra computational complexity), and SIC (with complexity $\mathcal{O}(ML|\mathcal{S}_s|S+|\mathcal{S}_s|^2LS)$) as in the MS-MRA case, except for employing the IISD block. In Step 1 of IISD, the complexity for computing $\mathbf{f}_i$ and implementing polar decoder are $\mathcal{O}(  (M n_c+M^2)T_{I}S\sum_{j=1}^J|\mathcal{D}_j| )$ and $\mathcal{O}(T_{I}n_c\log n_cS\sum_{j=1}^J|\mathcal{D}_j| )$, respectively, where $T_{I}$ denotes the number of iterations of IISD. In the Step 2 of IISD, computing $\hat{\mathbf{h}}_{i}$ and $e_k$ has the complexity of $\mathcal{O}(T_{I}(n_c+n_p)^2S\sum_{j=1}^J|\mathcal{D}_j|+T_{I}(n_c+n_p)MS\sum_{j=1}^J|\mathcal{D}_j|)$ and $\mathcal{O}(T_{I}(J-1)n_pMS\sum_{j=1}^J|\mathcal{D}_j|)$, respectively. The computational complexity of obtaining ${\hat{\bf h}}_i$ in Step 3 of IISD is $\mathcal{O}(T_{I}(L^2+LM)S\sum_{j=1}^J|\mathcal{D}_j|)$. Then, replacing $|\mathcal{D}_j|$ and $|\mathcal{S}_s|$ with their approximate values (discussed in the previous paragraph), the overall computational complexity of the MS-MRA-WOPBE is bounded by $\mathcal{O}\Bigl(n_p^2MJS+\max\left(\left((L^2+M^2)+ML\right)T_{I}JK_a,LK_a(M+K_a/S)\right)\Bigl)$. 

For comparison purposes, the dominant per-iteration computational complexity of the FASURA in \cite{Gkagkos2022FASURA} (which is due to energy detector and SIC operation) can also be computed as $\mathcal{O}\left(M(n_p+L^\prime n_c)2^{B_f}+K_a(nM+n^2) \right)$, where $B_f$ denotes the number of pilot bits, $n$ is the frame length, and $L^\prime$ is the length of the spreading sequence. 

\section{Numerical Results}
\label{section_numResults}
\begin{figure}[t!]
		\centering
	    \includegraphics[width=1.01\linewidth]{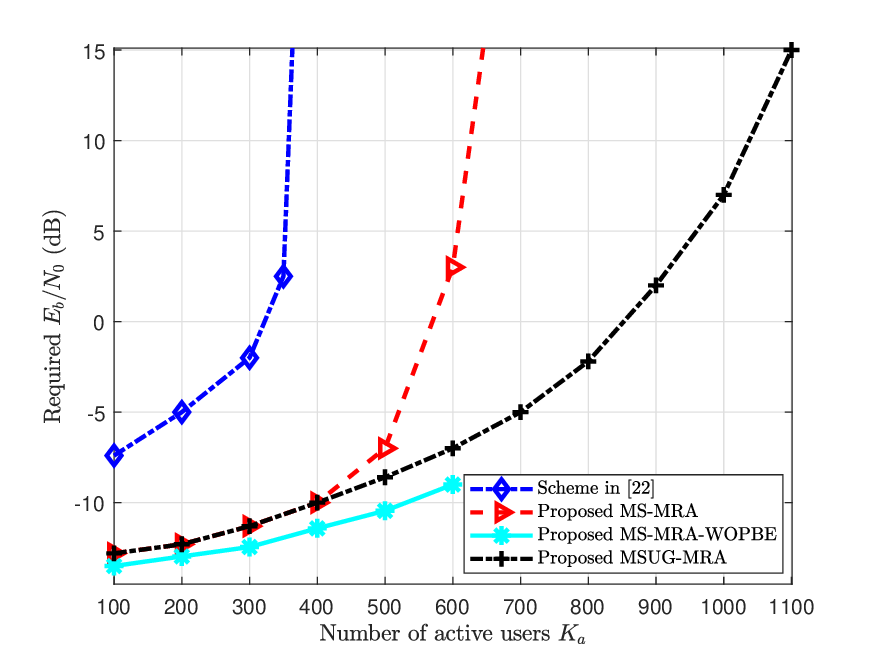}
		\caption{{\small The required $E_b/N_0$ in the proposed MIMO set-ups and the scheme in \cite{fengler2021non} for $L\approx 200$, $M=100$, and $P_e=0.05$.}	}
		\label{MIMO_short}
	\end{figure}

We provide a set of numerical results to assess the performance of the proposed URA set-ups. In all the results, we set $B = 100$, the number of CRC bits $r = 11$, the Neyman-Pearson threshold $\gamma = 0.1$, and the list size of the decoder to $64$. For MS-MRA and MSUG-MRA, we set the frame length $n\approx 3200$, and $P_e = 0.05$. The corresponding values for the MS-SRA and MSUG-SRA are $n\approx 30000$, and $P_e = 0.1$. \\
\indent In Fig. \ref{MIMO_short}, the performance of the proposed MS-MRA and MSUG-MRA is compared with the short blocklength scheme of \cite{fengler2021non} with the number of antennas $M=100$ and slot length $L= 200$. (In this scenario, we consider a fast-fading environment, where the coherence blocklength is considered as $L_c = 200$). To facilitate a fair comparison, we consider $(J,n_p,n_c) =  (2,32,128)$ ($L=192$) and $P_p/P_c =1$ ($\phi = 1$ for MSUG-MRA) for all the proposed schemes. For MSUG-MRA, the value of $G$ is set as $G=1$ for $K_a\leq 400$, $G = 3$ for $ K_a= 500$, $G = 6$ for $600\leq K_a \leq 800$, $G = 8$ for $900 \leq K_a\leq 1000$, and $G=10$ for $K_a>1000$. The superiority of the proposed schemes over the one in \cite{fengler2021non} is mostly due to the more powerful performance of the polar code compared to the simple coding scheme adopted in \cite{fengler2021non} and the use of the SIC block, which significantly diminishes the effect of interference. We also observe that MS-MRA-WOPBE outperforms MS-MRA, which is due to 1) employing IISD, which iteratively improves the accuracy of the channel estimation, and 2) lower coding rate by not encoding the pilot bits. Besides, the range of the number of active users that are detected by the MSUG-MRA is higher than those of MS-MRA and MS-MRA-WOPBE schemes. This improvement results from randomly dividing users into different groups, which provides each group with a lower number of active users (hence a lower effective interference level).

\begin{figure}[t!]
		\centering
\includegraphics[width=1.01\linewidth]{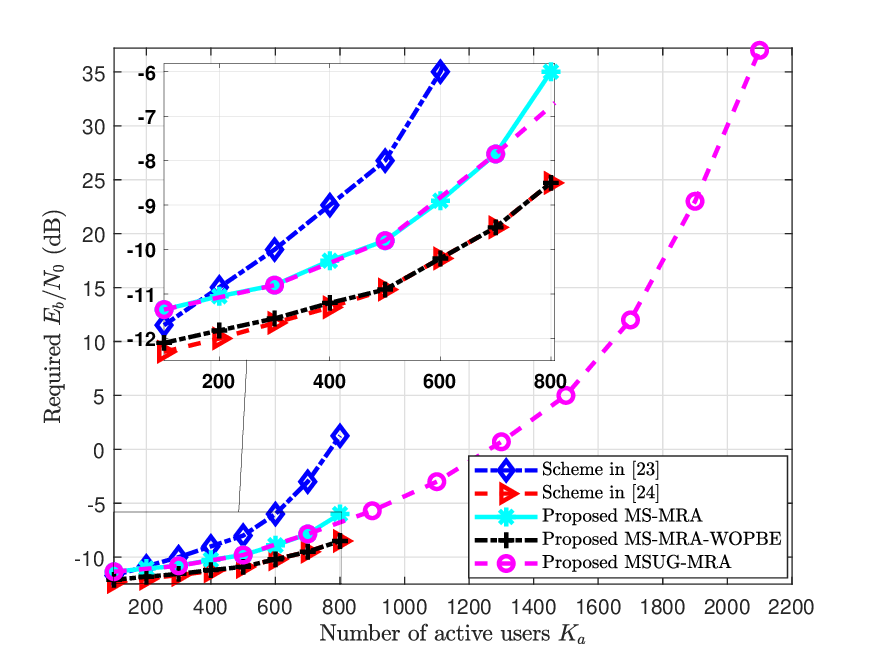}
		\caption{{\small The required $E_b/N_0$ in the proposed MIMO set-ups and the results in \cite{fengler2020pilot,Gkagkos2022FASURA} for $M=50$.}}	\label{MIMO_long}
	\end{figure}

\begin{figure}[t!]
	\centering
\includegraphics[width=1.01\linewidth]{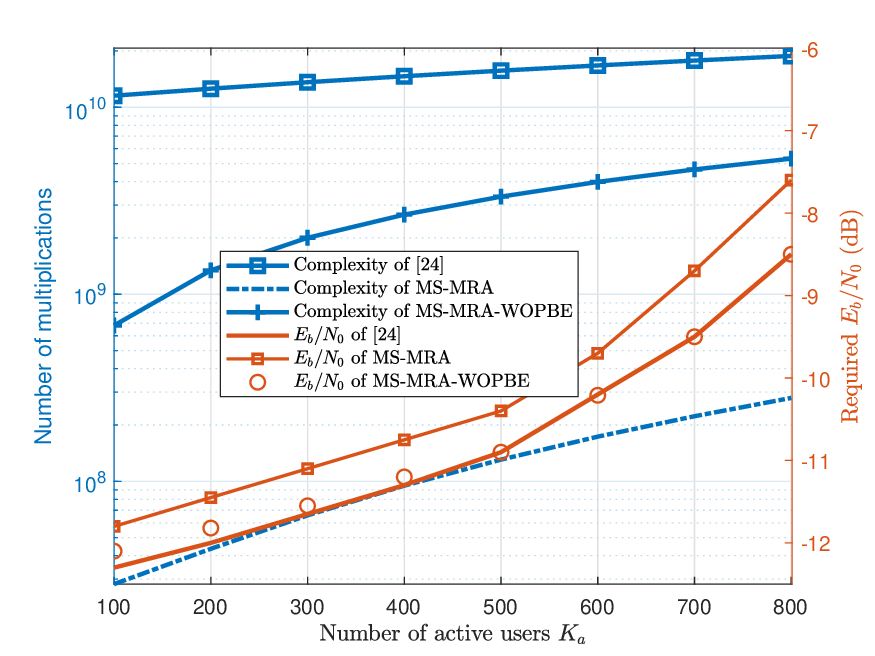}
		\caption{{\small Performance-complexity curve for the proposed MIMO schemes and FASURA in \cite{Gkagkos2022FASURA}.}	}	\label{Fig_perf_complxty}
	\end{figure}

 In Fig. \ref{MIMO_long}, we compare the proposed MS-MRA and MSUG-MRA with the ones in \cite{ fengler2020pilot, Gkagkos2022FASURA}, considering the slow-fading channel with coherence blocklength $L_c=3200$. We set $(J,n_p,n_c) = (2,256,512)$, $M=50$, $P_p/P_c =0.66$ for MS-MRA. We choose $(J,n_p,n_c, G) = (2,256,512, 1)$ for $K_a \leq 700$, $(J,n_p,n_c, G) = (2,64,512, 6)$ for $K_a = 900$, and $(J,n_p,n_c, G) = (2,64,512, 18)$ for $K_a > 900$ with $\phi = 0.66$. Thanks to employing the slotted structure, SIC, and orthogonal pilots, all the proposed schemes have superior performance compared to \cite{fengler2020pilot}. Due to employing random spreading and an efficient block called NOPICE, FASURA in \cite{Gkagkos2022FASURA} performs better than the proposed MS-MRA and MSUG-MRA in the low $K_a$ regimes; however, its performance is worse than the MSUG-MRA in higher values of $K_a$ (thanks to the random user grouping employed in MSUG-MRA). The proposed MS-MRA-WOPBE also shows a similar performance as FASURA. To achieve the result in Fig.~\ref{MIMO_long}, FASURA sets $n_p = 896$, $L^\prime = 9$, $n_c = 256$, $n = 3200$, $B_f = 16$, and $M=50$. The order of computational complexity for these schemes is given in the performance-complexity plot in Fig. \ref{Fig_perf_complxty}. It can be interpreted from this figure that the proposed MS-MRA-WOPBE has comparable accuracy to FASURA while offering a lower computational complexity. Note also that despite the higher required $E_b/N_0$ compared to FASURA, MS-MRA offers very large savings in terms of computational complexity, which is attributed to employing orthogonal pilots, slotted structure, and simpler decoding blocks.
 
 As a further note, FASURA considers $2^{B_p}$ possible spreading sequences of length $L^\prime$ for each symbol of the polar codeword; hence every transceiver should store $n_c 2^{B_p}$ vectors of length $L^\prime$, as well as a pilot codebook of size $2^J\times n_p$. For typical values reported in \cite{Gkagkos2022FASURA}, the BS and every user must store $1.6\times 10^7$ vectors of length $9$ and a matrix of size $5.8\times 10^7$. For the proposed schemes in this paper, every transceiver must store only an orthogonal codebook of size $n_p\times n_p$, where $n_p = 256$. Thus, FASURA requires about 3000 times larger memory than our proposed schemes, which may be restrictive for some target URA applications such as sensor networks, where a massive number of cheap sensors are deployed. Moreover, unlike FASURA, the proposed solutions are implementable with short blocklengths (see Fig. \ref{MIMO_short}), which makes them appropriate for fast fading scenarios as well.

\begin{figure}[t!]
	\centering
	    \includegraphics[width=1.01\linewidth]{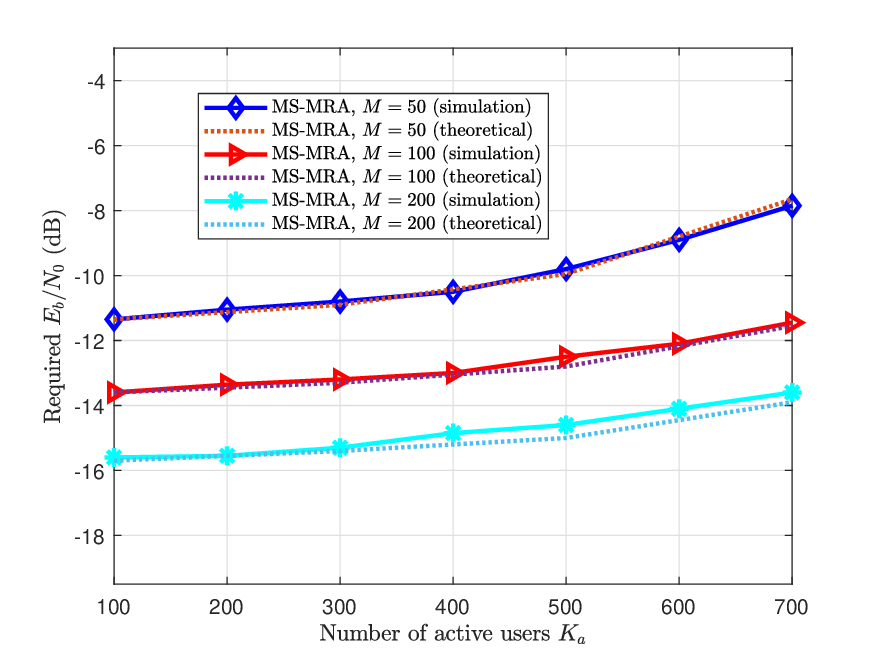}
		\caption{{\small Comparison of the simulation and analytical performance of the MS-MRA for different values of $M$.}	}	\label{fig_theory}
	\end{figure}

 In Fig. \ref{fig_theory}, we compare the theoretical PUPE in \eqref{eq18Ach} with the simulation results of the MS-MRA for  three different scenarios ($M=50, 100, 200$) with $P_p/P_c =0.66$ and $(J,n_p,n_c) = (2,256,512)$. It is shown that the approximate theoretical analysis well predicts the performance of the MS-MRA for $K_a\leq 700$, however, the results are not consistent for higher values of $K_a$. The reason for the mismatch for the $K_a> 800$ regime is the approximations employed while analyzing SIC in Lemma \ref{lemma_y_cy_p} (e.g., $n_c,n_p\gg 1$, $|\mathcal{S}_s|\gg 1$, uncorrelated QPSK codewords of two different users, and uncorrelated samples of $\mathbf{x}_i$).

Fig.~\ref{SISOPlot} compares the MS-SRA and MSUG-SRA with the existing single-antenna solutions \cite{kowshik2019energy,andreev2019low,andreev2020polar}. For both set-ups, we set $(J,n_p,n_c) = (2,64,512)$, $P_p/P_c = 1$ ($\phi = 1$ for MSUG-SRA), $(S,V) = (6,8)$ for $K_a\leq 200$, and $(S,V) = (12,4)$ for $K_a\geq 300$. For MSUG-SRA, we also choose $G=1$ for $K_a\leq 300$, $G=3$ for $500\leq K_a \leq 700$, and $G = 6$ for $K_a \geq 900$. It is observed that the proposed MS-SRA has a superior performance compared to the existing URA approaches for the low number of active users, However, it performs worse than the scheme in \cite{andreev2020polar} for higher values of $K_a$. Furthermore, the proposed MSUG-SRA outperforms existing solutions, and its effective range of $K_a$ is up to $1500$ users.

\section{Conclusions}
We propose a family of unsourced random access solutions for MIMO Rayleigh block fading channels. The proposed approaches employ a slotted structure with multiple stages of orthogonal pilots. The use of a slotted structure along with the orthogonal pilots leads to the lower computational complexity at the receiver, and also makes the proposed designs implementable for fast fading scenarios. We further improve the performance of the proposed solutions when the number of active users is very large by randomly dividing the users into different interleaver-power groups. The results show that the proposed MIMO URA designs are superior for both short and large blocklengths, while offering a lower computational complexity. 
\begin{appendices}
\section{Proof of Theorem \ref{Theorem_SINR}}
\label{Appendix_ch_dec}
\begin{mydef3}
\label{lemm_np_SL_infty}
Assuming that the transmitted data part contains uncorrelated and equally likely QPSK symbols, for $i, j \in \mathcal{S}_s$ and $n_p, n_c \to \infty$, the transmitted signals satisfy
\begin{align}
   \dfrac{1}{E_x}\mathbf{x}_{i}\mathbf{x}_{j}^H  &\overset{p}{\to} 0, \label{eq_xxH}
\end{align}
where $E_x= Jn_pP_p + n_c P_c$.
\end{mydef3}
\begin{proof}
Let $\mathbf{b}_{ji}$ and $\mathbf{b}_{jr}$ be the $j$th pilots of the $i$th and $r$th users, and $\mathbf{v}_i$ and $\mathbf{v}_r$ be the corresponding polar-coded and QPSK-modulated signals. Since $\mathbf{b}_{ji}$ and $\mathbf{b}_{jr}$ are randomly chosen rows of the Hadamard matrix, $\mathbf{b}_{ji}\mathbf{b}_{ji}^T = n_p$ with probability $\dfrac{1}{n_p}$, and it is zero with probability $1-\dfrac{1}{n_p}$. Besides, for $n_c\to \infty$, $v_{it}$ and $v_{rt}$ are zero-mean and uncorrelated, where $v_{it}=[\mathbf{v}_i]_{(:,t)}$. Therefore, 
\begin{align}
\nonumber    \lim_{n_p,n_c\to \infty} &\mathbb{P}\left(  {\dfrac{1}{E_x}}|\mathbf{x}_r\mathbf{x}_i^H|>0\right)\\\nonumber
= & \lim_{n_p,n_c\to \infty} \mathbb{P}\left( {\dfrac{1}{E_x}} \left|P_p \sum_{j=1}^J \mathbf{b}_{jr}\mathbf{b}_{ji}^H+\mathbf{v}_{r}\mathbf{v}_{i}^H\right|>0\right)\\\nonumber
\leq & \lim_{n_p,n_c\to \infty} \mathbb{P}\left( \dfrac{P_p}{{E_x}}  \sum_{j=1}^J \left|\mathbf{b}_{jr}\mathbf{b}_{ji}^H\right|+{\dfrac{1}{E_x}}\left|\mathbf{v}_{r}\mathbf{v}_{i}^H\right|>0\right)\\\nonumber
\nonumber
\leq &\lim_{n_p,n_c\to \infty}   \sum_{j=1}^J \mathbb{P}\left(\dfrac{P_p}{{E_x}}\left|\mathbf{b}_{jr}\mathbf{b}_{ji}^H\right|>0\right)\\\nonumber
\indent & \indent \indent \indent \indent \indent \indent \indent \indent \indent +\mathbb{P}\left({\dfrac{1}{E_x}}\left|\sum_{t=1}^{n_c}v_{rt}v_{it}^H\right|>0\right)\\\nonumber
\approx &\lim_{n_p,n_c\to \infty}   \dfrac{JP_p}{n_p{E_x}}+\mathbb{P}\left(\dfrac{n_c}{{E_x}}\left|\mathbb{E}\left\{v_{rt}v_{it}^H\right\}\right|>0\right)\\\nonumber
 \approx & \  0.
\end{align}
Note that, strictly speaking, the uncorrelated QPSK symbol assumption is not accurate for coded systems. Nevertheless, it is useful to obtain a good approximation of SINR, as we will show later.
\end{proof}
\begin{figure}[t!]
   \centering
    \includegraphics[width=1.01\linewidth]{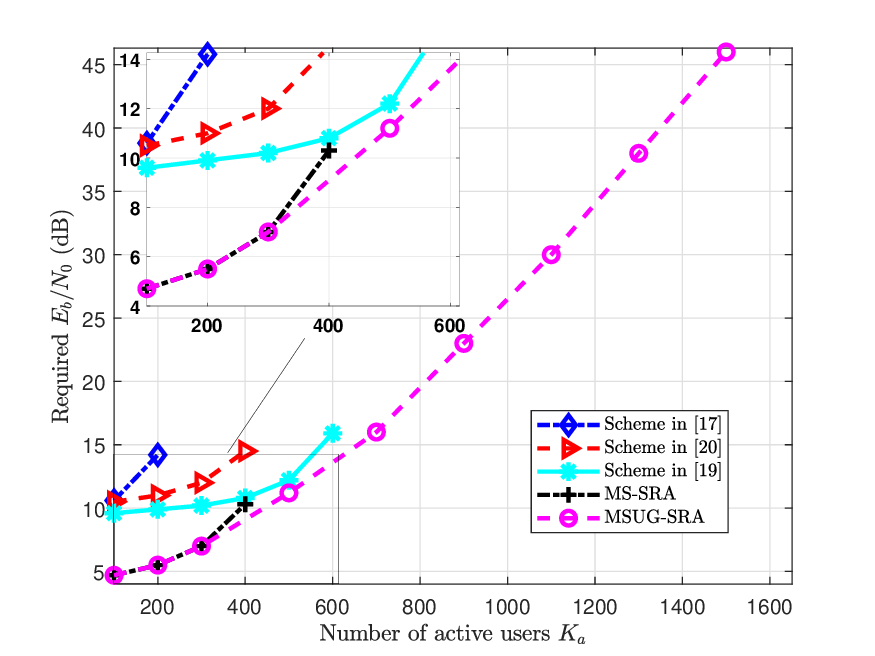}
    \caption{{\small The required $E_b/N_0$ of the proposed MS-SRA and MSUG-SRA for the case of single-antenna receiver}.}	\label{SISOPlot}
\end{figure}

\begin{mydef3}
\label{lemma_y_cy_p}
By applying LS-based SIC, the residual received signal matrices of pilot and coded parts can be written based on the signal and interference-plus-noise terms as
\begin{align}
        \mathbf{Y}^\prime_{p_j}\approx & \sqrt{P_p} \mathbf{h}_i\mathbf{b}_{ji}\mathbf{L}_{p_j}+\sqrt{P_p}\sum_{k\in \tilde{\mathcal{S}}_s, k\neq i}^{}\mathbf{h}_k\mathbf{b}_{jk}\mathbf{L}_{p_j}+ \mathbf{Z}_{n,p_j},\label{eq_pilot_received_resid}
\end{align}
\begin{align}
    \mathbf{Y}^\prime_{c} \approx  \mathbf{h}_i\mathbf{v}_{i}\mathbf{L}_{c}+\sum_{k\in \tilde{\mathcal{S}}_s, k\neq i}^{}\mathbf{h}_k\mathbf{v}_{k}\mathbf{L}_{c}+ \mathbf{Z}_{n,c},\label{eq_data_received_resid}       
\end{align}
where $\mathbf{h}_i \in \mathbb{C}^{M\times 1}$ is the channel coefficient vector of the $i$th user, $\mathbf{L}_{p_j}=\omega_{p_s}\mathbf{I}_{n_p}$, $\mathbf{L}_{c}=\omega_{c_s}\mathbf{I}_{n_c}$, and the elements of $\mathbf{Z}_{n,p_j}$ and $\mathbf{Z}_{n,c}$ are drawn from $\mathcal{CN}\left(0,\omega_{c_s}\sigma_z^2\right)$ and $\mathcal{CN}\left(0,\omega_{p_s}\sigma_z^2\right)$, respectively, with $\omega_{p_s}$ and $\omega_{c_s}$ are as defined in the statement of the Theorem \ref{Theorem_SINR}.

\end{mydef3}
\begin{proof}
Plugging \eqref{eq21} and \eqref{ch_esimation_sic} into \eqref{eq16}, we obtain
\begin{align}
\nonumber
\mathbf{Y^\prime} &=  \mathbf{H}_{\mathcal{S}_s}\mathbf{X}_{\mathcal{S}_s}\mathbf{L}+\mathbf{H}_{\tilde{\mathcal{S}}_s}\mathbf{X}_{\tilde{\mathcal{S}}_s}\mathbf{L}+\mathbf{Z}_s\mathbf{L}\\ \nonumber
        &=\mathbf{H}_{\tilde{\mathcal{S}}_s}\mathbf{X}_{\tilde{\mathcal{S}}_s}\mathbf{L}+\mathbf{Z}_s\mathbf{L} \\
        &=  \mathbf{h}_i\mathbf{x}_{i}\mathbf{L}+ \sum_{k\in \tilde{\mathcal{S}}_s, k\neq i}^{}\mathbf{h}_k\mathbf{x}_{k}\mathbf{L}+\mathbf{Z}_n , \label{eq_19_receivedSIC}
\end{align}
where $\mathbf{L} = \mathbf{I}_L-\mathbf{X}_{\mathcal{S}_s}^H(\mathbf{X}_{\mathcal{S}_s}\mathbf{X}_{\mathcal{S}_s}^H)^{-1} \mathbf{X}_{\mathcal{S}_s}$, and $\mathbf{Z}_n=\mathbf{Z}_s \mathbf{L}$.  Since $\mathbf{L}^H \mathbf{L}=\mathbf{L}$ and $\mathbf{Z}_{s}  \sim \mathcal{CN}\left(\mathbf{0},\sigma_z^2 \mathbf{I}_L\right)$, we have 
\begin{align}
\mathbf{Z}_{n}  \sim \mathcal{CN}\left(\mathbf{0},\sigma_z^2\mathbb{E}\{\mathbf{L}\}\right) \label{eq_zn}    .
\end{align}
 Since the values of $n_p$ and $n_c$ are large, and using \eqref{eq_xxH}, we have $\dfrac{1}{{E_x}}\mathbf{X}_{\mathcal{S}_s}\mathbf{X}_{\mathcal{S}_s}^H \approx  \mathbf{I}_{|\mathcal{S}_s|} $, where $E_x= Jn_pP_p + n_c P_c$. In other words, we can approximate $\mathbf{L}$ as 
\begin{align}
\mathbf{L} \approx   \mathbf{I}_L-\dfrac{1}{E_x}\sum_{r\in \mathcal{S}_s}^{}\mathbf{x}_{r}^H\mathbf{x}_{r} \label{eq_approx_L1}.
\end{align}
Using the weak law of large numbers, and assuming samples of $\mathbf{x}_{r}$ to be uncorrelated and $|\mathcal{S}_l|\gg 1$, we can rewrite $\mathbf{L}$ in \eqref{eq_approx_L1} as
\begin{align}
    \mathbf{L}\approx \begin{bmatrix}
    \mathbf{L}_{p_1} &... & \mathbf{0} & \mathbf{0} \\
    \vdots &\ddots &\vdots&\vdots \\
      \mathbf{0} &... & \mathbf{L}_{p_J} & \mathbf{0} \\
    \mathbf{0}&... &   \mathbf{0} & \mathbf{L}_{c}
    \end{bmatrix} ,\label{eq_xx}
\end{align}
where $\mathbf{L}_{p_j}=\omega_{p_s}\mathbf{I}_{n_p}$ and $\mathbf{L}_{c}=\omega_{c_s}\mathbf{I}_{n_c}$ with  $\omega_{p_s}=\omega_{c_s}=1-\dfrac{| \mathcal{S}_s|}{L}$ if the transmitted signals are randomly interleaved, and $\omega_{p_s}=1-\dfrac{1}{E_x}P_p| \mathcal{S}_s|$, $\omega_{c_s}=1-\dfrac{1}{E_x}P_c| \mathcal{S}_s|$, otherwise.

Letting $\mathbf{Z}_{n}=\left[\mathbf{Z}_{n,p_1}, \hdots, \mathbf{Z}_{n,p_J}, \mathbf{Z}_{n,c}\right]$, we can infer from \eqref{eq_zn} and \eqref{eq_xx} that the elements of $\mathbf{Z}_{n,p_j}$ and $\mathbf{Z}_{n,c}$ approximately follow $\mathcal{CN}\left(0,\omega_{p_s}\sigma_z^2\right)$ and $\mathcal{CN}\left(0,\omega_{c_s}\sigma_z^2\right)$, respectively. Besides, using \eqref{eq_xx} and the signal structure in \eqref{eq10_old}, we can divide \eqref{eq_19_receivedSIC} into pilot and coded parts as in \eqref{eq_pilot_received_resid} and \eqref{eq_data_received_resid}.
\end{proof}
\begin{mydef3}
\label{lemma1}
The estimated channel coefficients of a non-colliding user approximately satisfy the following expressions:
\begin{subequations}\label{eq_51_1}
 \begin{align}
\mathbb{E}\{\| \hat{\mathbf{h}}_{i}\|^2\} &\approx \omega_{p
_s}^2 \mathbb{E}\{\|\mathbf{h}_i\|^2\}+\dfrac{M \omega_{p_s}\sigma_z^2}{ n_{p}P_{p}}, \\
\mathbb{E}\{ |\hat{\mathbf{h}}_{i}^H\mathbf{h}_{k}|^2 \}&\approx \omega_{p_s}^2 \mathbb{E}\{\|\mathbf{h}_i\|^2\}+\dfrac{M \omega_{p_s}\sigma_z^2}{ n_{p}P_{p}},\\
\mathbb{E}\{ |\hat{\mathbf{h}}_{i}^H\mathbf{h}_{i} |^2\} &\approx  \omega_{p_s}^2\mathbb{E}\{\|\mathbf{h}_i\|^4\} +\dfrac{\omega_{p_s}\sigma_z^2}{ n_{p}P_{p}} \mathbb{E}\{\|\mathbf{h}_i\|^2\} .
\end{align}
\end{subequations}
\end{mydef3}
\begin{proof}
Using the approximation of $\mathbf{Y}^\prime_{p_j}$ in \eqref{eq_pilot_received_resid} in \eqref{eq10_Chestimate}, the channel coefficient vector of the $i$th user can be estimated as
\begin{align}
 \nonumber
     \hat{\mathbf{h}}_{i}  \approx &\ \dfrac{\omega_{p_s}}{ n_{p}} \mathbf{h}_i\mathbf{b}_{ji}\tilde{\mathbf{b}}_{jk}^H+\dfrac{\omega_{p_s}}{ n_{p}}\sum_{f\in \tilde{\mathcal{S}}_s, f\neq i}^{}\mathbf{h}_f\mathbf{b}_{jf}\tilde{\mathbf{b}}_{jk}^H+  \mathbf{z}_{p_j,n}\\
    \numeq{a} & \ \omega_{p_s}\mathbf{h}_i+\mathbf{z}_{p_j,n},\label{eq62}
\end{align}
where $\mathbf{z}_{p_j,n} = \dfrac{1}{ n_{p}\sqrt{P_{p}}}\mathbf{Z}_{n,p_j}\tilde{\mathbf{b}}_{jk}^H$, and in (a), we use the assumption that the $i$th user is non-colliding, hence $\tilde{\mathbf{b}}_{jk}$ is only selected by the $i$th user ($\mathbf{b}_{ji}=\tilde{\mathbf{b}}_{jk}$ and  $\mathbf{b}_{jf}\neq  \tilde{\mathbf{b}}_{jk}^H$ for $f\in \tilde{\mathcal{S}}_s, f\neq i$). We can argue the following approximation $\mathbf{z}_{p_j,n}  \sim \mathcal{CN}\left(0,\dfrac{\omega_{p_s}\sigma_z^2}{ n_{p}P_{p}}\right)$. Using~\eqref{eq62}, we can show that $\mathbb{E}\{\| \hat{\mathbf{h}}_{i}\|^2\} \approx \omega_{p_s}^2 \mathbb{E}\{\|\mathbf{h}_i\|^2\}+\dfrac{M \omega_{p_s}\sigma_z^2}{ n_{p}P_{p}}$, $\mathbb{E}\{ |\hat{\mathbf{h}}_{i}^H\mathbf{h}_{i} |^2\} \approx \omega_{p_s}^2\mathbb{E}\{\|\mathbf{h}_i\|^4\} +\dfrac{\omega_{p_s}\sigma_z^2}{ n_{p}P_{p}} \mathbb{E}\{\|\mathbf{h}_i\|^2\}$, and $\mathbb{E}\{ |\hat{\mathbf{h}}_{i}^H\mathbf{h}_{k}|^2 \} = \mathbb{E}\{\| \hat{\mathbf{h}}_{i}\|^2\}$.
\end{proof}
Plugging \eqref{eq_data_received_resid} into the MRC expression in \eqref{eq17}, $\hat{\mathbf{v}}_{i} $ can be estimated as 
\begin{align}
    \hat{\mathbf{v}}_{i} 
\approx   \omega_{c_s} \hat{\mathbf{h}}_{i}^H\mathbf{h}_i \mathbf{v}_i+ \mathbf{z}_{in},\label{eq_vprime}
\end{align}
where the first term on the right-hand side is the signal term, and $\mathbf{z}_{in}=\sum_{k\in \tilde{\mathcal{S}}_s, k\neq i}^{} \hat{\mathbf{h}}_{i}^H\mathbf{h}_k\mathbf{v}_{k}\mathbf{L}_{c}+ \hat{\mathbf{h}}_{i}^H\mathbf{Z}_{n,c}$ is the interference-plus-noise term. Since $\mathbf{L}^H\mathbf{L}=\mathbf{L}$, and using \eqref{eq_xx}, we can show $\mathbf{L}_c^H\mathbf{L}_c\approx \mathbf{L}_c$. Therefore, by employing Lemma \ref{lemma1}, we can approximate $\mathbf{z}_{in}\sim \mathcal{CN}(\mathbf{0},\sigma_{in}^2\mathbf{I}_{n_c})$, where 
\begin{align}
\nonumber
    \sigma_{in}^2= \omega_{c_s}\left(P_c  (|\tilde{\mathcal{S}}_s|-1) +\sigma_z^2\right)\left(\omega_{p_s}^2 \mathbb{E}\{\|\mathbf{h}_i\|^2\}+\dfrac{M\omega_{p_s} \sigma_z^2}{ n_{p}P_{p}}\right).   
\end{align}
Besides, the per-symbol power of the signal term can be obtained as $\sigma_s^2 \approx \omega_{c_s}^2 \mathbb{E}\{|\hat{\mathbf{h}}_{i}^H\mathbf{h}_i|^2\}P_c$. Then, using Lemma \ref{lemma1}, the SINR of $\hat{\mathbf{v}}_{i}$ can be calculated as in \eqref{sinr_initial}.
\section{Proof of Theorem \ref{Thm_collision}}
\label{Appendix_collision}
In the first iteration of the $s$th slot, since $K_s$ users have selected one out of $n_p$ pilots randomly, the number of users that select an arbitrary pilot approximately follows a Poisson distribution with the parameter $K_s/n_p$. In the $k$th iteration of the $s$th slot, let $T_{j,i}^{(k)}$ be the average number of $i$-collision pilots (pilots selected by $i$ different users) in the $j$th pilot part. We have 
\begin{align}
     T_{j,i}^{(1)}\approx n_p f_p(i;K_s/n_p),\label{Eq_numCol}
\end{align}    
where $f_p(i;a)$ denotes the PMF of the Poisson distribution with the parameter $a$. The average number of i-collision users in the $k$th iteration of the $j$th pilot part is then calculated as $K_{j,i}^{(k)}\approx i T_{j,i}^{(k)}$. Supposing that in the $k$th iteration (using the assumption in \eqref{assumption1}), the decoder employs the $j$th pilot part for channel estimation, the removed user is non-colliding (1-collision) in its $j$th pilot part (we assume that the decoder can only decode the non-colliding users), and it is in $i$-collision in its $j^\prime$th ($j^\prime\neq j$) pilot part with probability $p_{i,j^\prime}^{(k)} = \dfrac{K_{j^\prime,i}^{(k)}}{K_s-k+1}$. Therefore, removing a user from the $j$th pilot part results in

\begin{itemize}
    \item In the $j$th pilot part, we have $T_{j,1}^{(k+1)}=T_{j,1}^{(k)}-1$, and $T_{j,i}^{(k+1)}=T_{j,i}^{(k)}$ for $i>1$.
    \item In the $j^\prime$th pilot part ($j^\prime\neq j$), we have $T_{j^\prime,i}^{(k+1)}=T_{j^\prime,i}^{(k)}+p_{i+1,j^\prime}^{(k)}-p_{i,j^\prime}^{(k)}$.
\end{itemize}
The collision probability of the $j$th pilot part in the $t$th iteration is then obtained as $P_{col}(j,t) = 1- \dfrac{T_{j,1}^{(t)}}{K_s-t+1}$. Finally, by approximating $T_{j,i}^{(t)}$ by its average over different pilot parts (i.e., $T_{j,i}^{(t)}\approx N_i^{(t)}:=\dfrac{1}{J}\sum_{j=1}^J T_{j,i}^{(t)}$) in above equations, the results in Theorem \ref{Thm_collision} are obtained. Note that since all the pilot parts are equally likely in the first iteration, we have $N_i^{(1)}\approx T_{j,i}^{(1)}\approx n_p f_p(i;K_s/n_p), \forall j=1,...,J$.
\end{appendices}

\end{document}